\documentclass[apj]{emulateapj}
\usepackage{apjfonts}
\newcommand{\lya}{\mbox{Ly$\alpha$}}

\newcommand{\lobs}{\mbox{$L_\mathrm{Ly\alpha}^\mathrm{obs}$}}
\newcommand{\fesclya}{\mbox{$f_\mathrm{esc}^\mathrm{Ly\alpha}$}}
\newcommand{\fesclyc}{\mbox{$f_\mathrm{esc}^\mathrm{LyC}$}}

\newcommand{\tlya}{\mbox{$T_\mathrm{Ly\alpha}^\mathrm{IGM}$}}

\newcommand{\lth}{\mbox{$L_\mathrm{Ly\alpha}^\mathrm{th}$}}
\newcommand{\ewth}{\mbox{$\mathrm{EW_{Ly\alpha}^{th}}$}}
\newcommand{\ewlya}{\mbox{$\mathrm{EW_{\rm Ly\alpha}}$}} 
\newcommand{\ewint}{\mbox{$\mathrm{EW_{Ly\alpha}^{int}}$}} 
 
\newcommand{\nz}{\mbox{$N_\mathrm{cold}Z_\mathrm{cold}$}}

\newcommand{\taucont}{\mbox{$\tau_d^\mathrm{c}$}}
\newcommand{\taudlya}{\mbox{$\tau_d^\mathrm{Ly\alpha}$}}
\newcommand{\taudclya}{\mbox{$\tau_d^\mathrm{c}(\lambda_\mathrm{Ly\alpha})$}}

\newcommand{\unitlum}{\mbox{$h^{-2}~\mathrm{ergs~s^{-1}}$}}

\shortauthors{
Kobayashi,
Totani,
\& Nagashima
}

\shorttitle{
UV LF AND EW DISTRIBUTION OF Ly$\alpha$ EMITTERS
}

\begin{document}
 
\title{
Lyman Alpha Emitters in Hierarchical Galaxy Formation II.
UV Continuum Luminosity Function and Equivalent Width
Distribution
}

\author{Masakazu A.R. Kobayashi}
\affil{Optical and Infrared Astronomy Division, National Astronomical
Observatory of Japan, Mitaka, Tokyo 181-8588, JAPAN}

\author{Tomonori Totani}
\affil{Department of Astronomy, School of Science, Kyoto University,
Sakyo-ku, Kyoto 606-8502, JAPAN}

\and
 
\author{Masahiro Nagashima} 
\affil{Faculty of Education, Nagasaki University, Nagasaki, 852-8521, JAPAN}
  
\email{mark.kobayashi@nao.ac.jp}

%%%%%%%%%%%%%%%%%%%%%%%%%%%%%%%%%%%%%%%%%%%%%%%%%%%%%%%%%%%
% The abstract should be a single paragraph of            %
%  not more than 250 words, and should not contain        %
%  reference citations.                                   %
%%%%%%%%%%%%%%%%%%%%%%%%%%%%%%%%%%%%%%%%%%%%%%%%%%%%%%%%%%%
\begin{abstract}

  We present theoretical predictions of UV continuum luminosity
  function (UV LF) and $\lya$ equivalent width (EW) distribution of
  Lyman alpha emitters (LAEs) in the framework of the hierarchical
  clustering model of galaxy formation.  The model parameters about
  LAEs were determined by fitting to the observed $\lya$ LF at $z=5.7$
  in our previous study, and the fit indicates that extinction of
  $\lya$ photons by dust is significantly less effective than that of
  UV continuum photons, implying clumpy dust distribution in
  interstellar medium.  We then compare the predictions about UV LFs
  and EW distributions with a variety of observations at $z \sim
  $3--6, allowing no more free parameters and paying careful attention
  to the selection conditions of LAEs in each survey.  We find that
  the predicted UV LFs and EW distributions are in nice agreement with
  observed data, and especially, our model naturally reproduces the
  existence of large EW LAEs ($\gtrsim 240$~\AA) without introducing
  Pop III stars or top-heavy initial mass function.  We show that both
  the stellar population (young age and low metallicity) and
  extinction by clumpy dust are the keys to reproduce large EW LAEs.
  The evidence of EW enhancement by clumpy dust is further
  strengthened by the quantitative agreement between our model and
  recent observations about a positive correlation between EW and
  extinction.  The observed trend that brighter LAEs in UV continuum
  tend to have smaller mean EW is also reproduced, and the clumpy dust
  is playing an important role again for this trend.  We suggested in
  our previous study that the transmission of intergalactic medium for
  $\lya$ emission rapidly decreases from $z \sim 6$ to 7 by the
  fitting to $\lya$ LFs, and this evidence is quantitatively
  strengthened by the comparison with the UV LF and EW distribution at
  $z \sim 6.6$.

\end{abstract}
 
%%%%%%%%%%%%%%%%%%%%%%%%%%%%%%%%%%%%%%%%%%%%%%%%%%%%%%%%%%%%
% The subject headings (a maximum of six) should be listed %
%  after the abstract.                                     %
% The current list of subject headings is printed in       %
%  the Annual Index to the Journal and is available online %
%  at http://www.journals.uchicago.edu/ApJ/keywords.html.  %
%%%%%%%%%%%%%%%%%%%%%%%%%%%%%%%%%%%%%%%%%%%%%%%%%%%%%%%%%%%%
\keywords{
galaxies: evolution ---
galaxies: formation ---
galaxies: high-redshift ---
methods: numerical
}

%%%%%%%%%%%%%%%%%%%%%%%%%%%%%%%%%%%%%%%%%%%%%%%%%%%%%%%%%%%%%%%%%%%%%

 \section{INTRODUCTION}
 \label{section:intro}

 Detecting redshifted $\lya$ emission with narrow-band imaging is a
 powerful strategy to seek for galaxies in the early universe.
 Indeed, many galaxies have been detected through this method in the
 last decade (e.g., Cowie \& Hu 1998; Rhoads et al. 2000; Taniguchi et
 al. 2005; Murayama et al. 2007; Ouchi et al. 2008), and they are
 called $\lya$ emitters (LAEs).\footnote{In this paper, so-called
 $\lya$ blobs that extend on super-galactic scale are not considered,
 and they are treated as a separate population from LAEs that we focus
 on.} Because of their strong $\lya$ emission lines, LAEs can be
 detected even at very high redshifts, as demonstrated by the fact
 that the spectroscopically confirmed highest-redshift galaxy so far
 ($z = 6.96$) was found by this method (Iye et al. 2006; Ota et
 al. 2008).  Moreover, they are an invaluable population to probe the
 cosmic reionization history because the strength and profile of
 $\lya$ emissions from LAEs could significantly be affected because of
 absorption by neutral hydrogen in the intergalactic medium (IGM;
 e.g., Malhotra \& Rhoads 2004; Santos 2004; Haiman \& Cen 2005;
 Dijkstra et al. 2007a, b; Dayal et al. 2008; Mesinger \& Furlanetto
 2008).

 The physical properties (such as mass, age or metallicity) of LAEs
 and the connection with other high-$z$ galaxy population [e.g.,
 Lyman-break galaxies (LBGs)] have been poorly understood because of
 the faintness of their continua.  On the other hand, their
 statistical properties such as luminosity functions (LFs) in terms of
 $\lya$ emission ($\lya$ LF) and rest-frame UV continuum luminosities
 (UV LF), and $\lya$ equivalent width (EW) distributions, have more
 firmly been established because of the increase of survey fields and
 available samples by different authors (Hu et al. 2004; Kashikawa
 et al. 2006; Shimasaku et al. 2006; Dawson et al. 2007; Gronwall et
 al. 2007; Murayama et al. 2007; Ouchi et al. 2008).

 In our previous study (Kobayashi et al. 2007, hereafter KTN07), we
 have constructed a new theoretical model for the $\lya$ LF of LAEs in
 the framework of hierarchical galaxy formation.  It is based on one
 of the latest semi-analytic model for galaxy formation, the Mitaka
 model (Nagashima \& Yoshii 2004; see also Nagashima et al. 2005), in
 which galaxies are formed based on the standard structure formation
 theory driven by cold dark matter.  There are several theoretical
 models for $\lya$ LF of LAEs based on analytic models (e.g., Mao et
 al. 2007; Dayal et al. 2008; Samui et al. 2009), semi-analytic galaxy
 formation models (e.g., Le Delliou et al. 2006; Orsi et al. 2008) or
 cosmological hydrodynamic simulations (e.g., Barton et al. 2004;
 Nagamine et al. 2008; Dayal et al. 2009).  The intrinsic
 production rate of $\lya$ photons within a galaxy is expected to be
 proportional to ionizing luminosity and hence it can be calculated if
 the star formation history (SFH) is known.  Therefore the escape
 fraction of $\lya$ photons from the galaxy is the key to predict LAE
 statistics.  However, the escape fraction is difficult to predict
 theoretically for a realistic configuration of interstellar gas and
 dust, although there are some theoretical studies to predict it
 through radiative transfer models in a simplified geometry of
 interstellar medium (ISM) and dust (e.g., Hansen \& Oh 2005; Verhamme
 et al. 2008).  Hence, most models simply assume a constant escape
 fraction, which clearly contradicts the recent observational results
 from local star-forming galaxies (e.g., Atek et al. 2008; \"{O}stlin
 et al. 2009).  The KTN07 model is unique because the two physical
 effects are incorporated in calculating the escape fraction:
 extinction by interstellar dust but with an amount that is different
 from that for UV continuum, and galaxy-scale outflows induced as
 supernova feedbacks.  In KTN07, we have shown that our outflow$+$dust
 model reproduces the observational $\lya$ LFs of the LAEs at $z\simeq
 3-6$.

 UV LFs and distributions of $\lya$ EW also provide important
 statistical information for LAEs in addition to $\lya$ LFs.  In
 particular, because $\lya$ EW is highly sensitive to SFH, age, and
 metallicity as well as initial mass function (IMF), the observed EW
 distribution can provide an opportunity to test the validity of the
 LAE models.  The purpose of this paper is to present a detailed and
 comprehensive comparison between the KTN07 model and the available
 observations of these statistical quantities at various redshifts.
 We will then focus on the two interesting issues described below from
 our analysis.
 
 The first issue is the existence of very large $\lya$ EW LAEs.
 Theoretical models of stellar evolution predict that the maximum
 $\lya$ EW powered by star-formation activity with the Salpeter IMF
 and the solar metallicity is 240~{\AA} (e.g., Charlot \& Fall 1993;
 Schaerer 2003).  However, observations have revealed that some
 fractions of LAE candidates have higher $\lya$ EWs than the maximum
 (e.g., Malhotra \& Rhoads 2002; Shimasaku et al. 2006; Dawson et
 al. 2007; Gronwall et al. 2007; Ouchi et al. 2008).  Although the
 existence of such high-EW LAEs has not yet firmly been confirmed
 because of faint UV continuum flux with large flux errors, such large
 EW LAEs could have significant implications for galaxy formation
 theory.  In fact, these observational results have often been used to
 argue that some fraction of metal-free stars (so-called Pop III
 stars) and/or a top-heavy IMF are required (e.g., Dijkstra \& Wyithe
 2007).  Another possibility is the enhancement of EW because of
 selective extinction for continuum photons by dust in dense clouds
 (Neufeld 1991; Hansen \& Oh 2006), and recent observations give some
 supports to this interpretation (Finkelstein et al. 2008, 2009a,
 2009b).  Since our model incorporates the effect of extinction by
 dust, it is interesting to see whether such an effect can
 quantitatively explain the large EW LAEs in our model.
 
 The second issue is the trend of LAE EW distributions as a function
 of UV luminosity.  In the rest-frame UV absolute magnitude
 $M_\mathrm{UV}$ versus $\ewlya$ plane, there is a lack of large EW
 LAEs with large UV luminosities, and the maximum EW systematically
 decreases with increasing UV luminosity (Shimasaku et al. 2006;
 Stanway et al. 2007; Deharveng et al. 2008; Ouchi et al. 2008).  This
 trend in the $M(\mathrm{UV})$-$\ewlya$ plane was first reported for
 LBGs at $z\sim 5-6$ by Ando et al. (2006), and we here call this
 feature as the Ando effect.  The physical origin of the Ando effect
 is not well understood.  A similar trend has already been known for
 broad emission lines of active galactic nuclei, which is called as
 the Baldwin effect (Baldwin 1977), but the physical origin can be
 completely different for the Ando effect on galaxies.  Ando et
 al. (2006) suggested that this relation may be attributed to higher
 metallicities in the UV-bright galaxies.  Another possible
 interpretation proposed by Ouchi et al. (2008) is that the average
 stellar population of the UV-bright galaxies is older than that of
 the UV-faint galaxies.  Schaerer \& Verhamme (2008) and Verhamme et
 al. (2008) also provided some qualitative predictions to the origin
 of the Ando effect based on the results from their radiative transfer
 model, but it was not examined quantitatively whether their
 prediction reproduces observational distribution in
 $M(\mathrm{UV})$-$\ewlya$ plane.  On the other hand, if the
 extinction effect by dust is important for the large EW LAEs as
 mentioned above, extinction should also be relevant to the Ando
 effect.  We will try to give a new theoretical explanation for the
 Ando effect based on our model.
 
 KTN07 found that, although the $\lya$ LFs of LAEs can be reproduced
 by the model at $z \sim 3$--6, the model overproduces $\lya$ LFs
 compared with observations at $z \gtrsim 6$, because of the rather
 sudden decrease of the observed $\lya$ LFs at $z \gtrsim 6$.  KTN07
 suggested that this can be interpreted by a rapid increase of IGM
 opacity against $\lya$ photons at $z \gtrsim 6$, giving an
 interesting implication for the cosmic reionization. Here we revisit
 this issue in terms of the increased statistical quantities of UV LFs
 and EW distributions, and examine the KTN07's interpretation.
 
 The paper will be organized as follows: In \S~\ref{section:model}, we
 describe our theoretical model, especially for the UV luminosity and
 EW calculation.  We compare the model results with the observed LAE
 statistical quantities at various redshifts $3\lesssim z \lesssim 6$
 in \S~\ref{section:comparison} and discuss about the above two
 issues.  After some discussions including implications for
 reionization (\S~\ref{section:discussion}), the summary will be given
 in \S~\ref{section:conclusions}. The background cosmology adopted in
 this paper is the standard $\Lambda$CDM model: $\Omega_M=0.3$,
 $\Omega_\Lambda=0.7$, $\Omega_b=0.04$, $h=0.7$, and $\sigma_8=0.9$.
 All magnitudes are expressed in the AB system, and all $\lya$ EW
 values in this paper are in the rest-frame.

%%%%%%%%%%%%%%%%%%%%%%%%%%%%%%%%%%%%%%%%%%%%%%%%%%%%%%%%%%%%%%%%%%%%%
 \section{MODEL DESCRIPTION}
 \label{section:model}

 A detailed description of our outflow$+$dust model about Ly$\alpha$
 emission from galaxies in the Mitaka model is given in KTN07.  Here
 we briefly summarize the essential treatments about the modeling of
 $\lya$ emission, and some changes and updates from the KTN07 model.

  \subsection{Ly$\alpha$ Photon Production} 
  \label{section:model-LyA-production}
  
  We consider only Ly$\alpha$ photons produced by star formation
  activity.  The Ly$\alpha$ photon production rate in a star-forming
  galaxy is calculated from ionizing UV photon ($\lambda < 912$~\AA)
  luminosity assuming that all the ionizing photons are absorbed by
  ionization of hydrogen within the galaxy, and $\lya$ photons are
  produced by the case B recombination.  The escape fraction of
  ionizing photons, $\fesclyc$, from a galaxy is not exactly zero in
  reality, but it is generally believed to be small (e.g., Inoue et
  al. 2006).  If $\fesclyc \ll 1$, the assumption of $\fesclyc = 0$ in
  this work is reasonable because the $\lya$ luminosity is
  proportional to $(1 - \fesclyc)$.

  The ionizing photon luminosity is calculated using the stellar
  evolution model of Schaerer (2003, hereafter S03) assuming the
  Salpeter IMF in 0.1--100~$M_\odot$. Its metallicity dependence is
  taken into account using the S03 model in a range of $Z/Z_{\odot} =
  0$--2.  The gas metallicity of each model galaxy is calculated in
  the Mitaka model.  In the quiescently star forming galaxies, which
  do not experience a major merger of galaxies, we relate the ionizing
  luminosity simply to star formation rate (SFR), because they are
  forming stars approximately constantly and the mean stellar age
  weighted by ionizing luminosity at the time of observation is much
  smaller than the time scale of SFR evolution.  However, in starburst
  galaxies triggered by major mergers of galaxies, SFR changes with a
  very short time scale and hence we exactly calculate ionizing photon
  luminosity by integrating the SFH.

  \subsection{Ly$\alpha$ Escape Fraction} 
  \label{section:model-LyA-escape}  

  The observed Ly$\alpha$ luminosity is determined by the escape
  fraction, \fesclya, of produced $\lya$ photons from their host
  galaxy.  It is difficult to predict $\fesclya$ of each model galaxy
  from the first principles in its realistic geometry of ISM and
  interstellar dust because of the resonant scattering of $\lya$
  photons.  The photon path could be very complicated depending on
  clumpiness and velocity fields of ISM, and extinction by
  interstellar dust for $\lya$ photons could be different from that
  for UV continuum photons.  We introduce a simple model where
  $\fesclya$ is separated into two parts:
  \begin{eqnarray}
   \fesclya = f_0 \ \frac{1 - \exp(- \taudlya)}{ \taudlya } \ ,
    \label{eq-fesclya}
  \end{eqnarray}
  where $f_0$ represents the reduction of escaping photons by physical
  effects not caused by interstellar dust, and the rest of the r.h.s.
  is for extinction by dust in the slab geometry [Rybicki \& Lightman
  1979, eq. (1.30); see also Clemens \& Alexander 2004, eqs.(1)-(3)].

  There are some physical effects that may reduce $f_0$ from the
  unity.  One such effect is too many times of resonant scatterings of
  $\lya$ photons.  When the scattering length is too small in some
  star forming regions in a galaxy, $\lya$ photons will diffuse out
  only by random-walk process, and the escape time scale can be
  extremely long, effectively resulting in a small $f_0$.  It should
  be noted that a possible deviation of $\fesclyc$ from the assumed
  value of $\fesclyc = 0$ is also absorbed within $f_0$.  Another
  effect on $f_0$ is the absorption by neutral hydrogen in IGM.
  $\lya$ photons that are blueshifted than the rest-frame of their
  host galaxy will significantly be attenuated by this effect at $z
  \gtrsim 3$, while those redshifted will escape freely.  We assume
  that this effect is independent of redshift at $z \lesssim 6$,
  because we expect that the evolution of absorption is not
  significant if the $\lya$ line profile is similar for LAEs at
  different redshifts.  However, at the very high redshifts of $z
  \gtrsim 6$ reaching the epoch of reionization, the damping wing
  effect of IGM absorption by the increase of IGM neutral fraction may
  become important, and in this case the damping wing would erase most
  of $\lya$ photons including those redshifted than the rest-frame of
  the host galaxy, resulting in a rapid evolution of IGM absorption at
  $z \gtrsim 6$ (e.g., Fan et al. 2006).  Therefore we treat the
  effect of damping wing separately from $f_0$, by introducing the IGM
  transmission $\tlya$, which is the fraction of $\lya$ photons
  transmitted after the absorption effect of the damping wing, as done
  in KTN07.  We emphasize that $\tlya$ represents only the effect of
  the damping wing at $z\gtrsim 6$, and hence $\tlya = 1$ at $z
  \lesssim 6$.  The modest effect of absorption by IGM at $z \lesssim
  6$ is effectively included in the parameter $f_0$.  The case of
  $\tlya < 1$ at $z \gtrsim 6$ will be discussed in
  \S~\ref{section:reionization}.

  The parameter $\taudlya$ is the effective optical depth of
  extinction by interstellar dust for $\lya$ photons.  Here, the slab
  type geometry has been assumed for the dust effect rather than the
  screen type geometry.  KTN07 found that the difference between the
  models assuming the slab or screen geometries is negligible about
  LAE $\lya$ LF predictions, and we here adopt the slab geometry
  because it has already been adopted for the dust extinction of
  continuum photons in the Mitaka model. In the Mitaka model, the dust
  optical depth for continuum photons are assumed to be proportional
  to the metal column density of cold gas, $N_{\rm cold} Z_{\rm
  cold}$, and we also assume this proportionality for the dust opacity
  for $\lya$ photons, $\taudlya$.  The metal column density can be
  calculated by the gas mass, galaxy size, and metallicity in the
  Mitaka model.  Therefore, $\taudlya$ is given by
  \begin{eqnarray}
   \taudlya = \frac{N_{\rm cold} Z_{\rm cold} }
    {(N_{\rm cold} Z_{\rm cold})_0^{\rm Ly\alpha}} \ ,
    \label{eq-taudlya}
  \end{eqnarray}
  with a model parameter $(N_{\rm cold} Z_{\rm cold})_0^{\rm
  Ly\alpha}$ that controls the strength of $\lya$ photon extinction.
  As mentioned above, this parameter can be different from that for UV
  continuum photons around 1216~{\AA}.  We determine this parameter by
  fitting to the observed $\lya$ LF of LAEs at $z=5.7$ (Shimasaku et
  al. 2006) independently of continuum extinction.
  
  We also consider galaxy-scale outflow by supernova feedback as a
  potential effect that could enhance the escape fraction because the
  velocity difference of ambient ISM produced by outflow may greatly
  increase the $\lya$ scattering length.  The outflow is considered
  only for starburst population triggered by major mergers of
  galaxies, and outflow occurs by supernova feedback.  We identify
  outflow-phase galaxies in a way consistent with the treatment of
  supernova feedback in the Mitaka model.  Hence, there are four
  phases of galaxies in our model: quiescent (non-starburst),
  pre-outflow starbursts, outflow starbursts, and post-outflow
  starbursts.  Although the distinction of model galaxies into these
  four phases seems somewhat arbitrary, we have done this based on
  physical considerations.  Briefly, the outflow phase onsets when the
  energy injected to ISM by supernova feedback exceeds the binding
  energy of ISM gas in the galaxy halo, and it continues during the
  dynamical time scale of halo ($\sim r_\mathrm{e} / V_\mathrm{c}
  \equiv t_\mathrm{esc}$, where $r_\mathrm{e}$ and $V_\mathrm{c}$ are
  the effective radius of galaxy and the circular velocity of its host
  halo, respectively).  The pre- and post-outflow starbursts are
  defined as galaxies before and after this outflow phase (see KTN07
  for more details).  The galactic wind terminates star formation, and
  then galaxies in the outflow or post-outflow phases are passively
  evolving.  We then introduce $f_0^{\rm wind}$ instead of
  eq.~(\ref{eq-fesclya}) as $\fesclya$ in the model galaxies under the
  outflow phase.  This is motivated by a physical consideration that
  $\fesclya$ becomes less sensitive to $\tau_d^{\rm Ly\alpha}$ in such
  outflowing condition because dust in ISM becomes sparse by the
  outflow and outflow drastically reduces the scattering optical depth
  of $\lya$ by neutral hydrogen in ISM.
  Finally, we assume that galaxies in post-outflow phase do not
  produce any $\lya$ emission, because ionizing luminosity should be
  reduced by the termination of star formation, and there is little
  amount of interstellar gas to absorb ionizing photons and produce
  $\lya$ photons within the galaxies.

  \subsection{LAE Model Parameter Determination} 
  \label{section:model-parameter}
  
  Consequently, there are three model parameters about LAEs: $(N_{\rm
  cold} Z_{\rm cold})_0^{\rm Ly\alpha}$, $f_0$, and $f_0^{\rm wind}$.
  We assume that they are independent of redshift, because these
  parameters reflect the physics about $\lya$ photon escape, that is
  independent of redshift.  These have been determined in KTN07 by the
  fit to the LAE $\lya$ luminosity function at $z=5.7$ measured by
  Shimasaku et al. (2006).  We found a unique set of the best-fit
  parameters\footnote{These values are slightly different from those
  listed in Table 1 of KTN07, because the dust geometry has been
  changed into the slab model in the dust+outflow model, and there was
  a tiny bug in ionizing luminosity calculation of the previous
  analysis. However, the overall agreement between the KTN07 LAE \lya
  \ LF model and the observations is not significantly affected.},
  which are: $\left(\nz \right)_0^{\rm Ly\alpha} =
  8.0^{+1.9}_{-1.4}\times 10^{21}~ [Z_\odot~ \mathrm{cm^{-2}}]$, $f_0
  = 0.23^{+0.02}_{-0.03}$ and $f_0^\mathrm{wind} = 0.36$.  The model
  parameters that are not related to LAEs in the Mitaka model are kept
  at the original values that have been determined by fits to
  observations of the local galaxies (Nagashima \& Yoshii 2004).  As
  demonstrated by KTN07, the predictions by this model are in overall
  agreement with observations of LAE $\lya$ LFs at various redshifts
  at $z \sim 3$--6.  It may be rather surprising, considering the
  complicated physics about $\lya$ emission, that this simple
  phenomenological model can explain observations with just three free
  parameters.  The outflow-phase escape fraction $f_0^{\rm wind}$ is
  not much different from $f_0$, and we get a reasonable agreement
  with observations even if we set $f_0^{\rm wind} = f_0$, indicating
  that the outflow effect is not significant about the $\lya$ escape
  fraction. This also means that the effective number of free
  parameters is further reduced, i.e., just two.
  
  We will then compare the UV continuum luminosities and EWs predicted
  by this model with observations.  It should be noted that all the
  model parameters have been determined by Nagashima \& Yoshii
  (2004) and KTN07, and there is no free parameter that can be
  adjusted to the new data compared in this work.  Therefore, the
  comparison with observations presented below provides an objective
  test for the validity of the KTN07 framework of LAE modeling.

  \subsection{Rest-Frame UV Luminosity}
  \label{subsec:model_UV}

  For the purpose of this paper, we need to calculate UV continuum
  luminosity of galaxies.  We present UV LFs at the rest-frame
  wavelength of $\lambda = $1500~{\AA}.  The unabsorbed (intrinsic) UV
  luminosity at this wavelength is calculated by using the S03 model
  in a similar way to the ionizing luminosity.  The observable UV
  luminosity is then calculated taking into account dust
  extinction. The amount of extinction magnitude $A_\lambda$ for
  continuum photons as a function of rest-frame wavelength $\lambda$
  has been calculated in the original Mitaka model, assuming that the
  optical depth $\tau_d^c(\lambda)$ is proportional to the metal
  column density of cold gas.  The wavelength dependence of
  $\tau_d^c(\lambda)$ is determined by the Galactic extinction curve
  (Pei 1992), and $A_\lambda$ is calculated from $\tau_d^c(\lambda)$
  assuming the slab type geometry, i.e.,
  \begin{eqnarray}
   10^{-0.4 A_\lambda} = \frac{1 - \exp[- \tau_d^c(\lambda)]}{
    \tau_d^c(\lambda) } \ .
  \end{eqnarray}
  The proportionality constant between $\tau_d^c$ and metal column
  density has been determined in the Mitaka model to fit the observed
  local galaxies (Nagashima \& Yoshii 2004).  

  For the starburst galaxies, we expect that their dust opacities are
  gradually reduced because SN explosions and subsequent galactic wind
  would heat and remove their interstellar cold gas.  It is difficult
  to predict analytically the amount of ISM left during the starburst
  activity.  Here we simply assume that ISM decreases exponentially
  with an e-folding time of $t_\mathrm{esc}$ around $t =
  t_\mathrm{wind}$.  We tested the sensitivity of our result to this
  prescription by using another model simply assuming no extinction in
  the outflow and post-outflow galaxies, and we confirmed that our
  main conclusions are not significantly affected.
  
  It is interesting to compare the extinction of UV continuum photons
  to that for $\lya$ photons in our model.  Both $\tau_d^{\rm
  Ly\alpha}$ and $\tau_d^c(\lambda)$ are assumed to be proportional to
  the metal column density of ISM, and the relation between the two
  around the \lya \ wavelength becomes
  \begin{equation}
   \taudlya \equiv q_d \, \taudclya,
    \label{eq-taudlya}
  \end{equation}
  where $q_d =0.149\pm 0.03$ is derived from the model parameters in
  our model.  We note here that $q_d$ is not a new free parameter in
  our model, because both $\taudlya$ and $\taudclya$ have been
  determined by the modeling described above.  The numerical constant
  $q_d$ in eq.~(\ref{eq-taudlya}) is called as the \textit{geometry
  parameter} or \textit{clumpiness parameter} introduced originally by
  Finkelstein et al.  (2008).\footnote{However, our definition of
  $q_d$ is slightly different from the original observational
  definition of $q_{\rm obs}$ by Finkelstein et al.  See
  \S~\ref{subsec:model_EW}.}  The value of $q_d$ effectively reflects
  the interstellar dust geometry.  The case of $q_d \gg 1$ means
  homogeneous ISM, in which the resonance scatterings make the photon
  path of $\lya$ photons much longer than that of UV continuum photons
  and hence $\lya$ photons suffer from much larger extinction.  On the
  other hand, the case of $q_d \ll 1$ is regarded as extremely clumpy
  ISM, in which $\lya$ photons are reflected by neutral hydrogen
  before entering the dense regions where a large amount of dust
  particles reside.  In such a case $q_d$ becomes smaller than unity,
  since $\lya$ photons propagate preferentially in low extinction
  regions, while UV continuum photons go through the dense dusty
  regions.  Detailed theoretical studies solving radiative transfer of
  $\lya$ photons in such clumpy ISM found that this is indeed possible
  (Neufeld 1991; Hansen \& Oh 2006). The value of $q_d =0.15$ derived
  from our best-fit model implies strong clumpiness of interstellar
  dust distributions in high-$z$ LAEs.

  \subsection{Rest-Frame $\lya$ Equivalent Width}
  \label{subsec:model_EW}
  
\begin{figure}
 \epsscale{1.}  
 \plotone{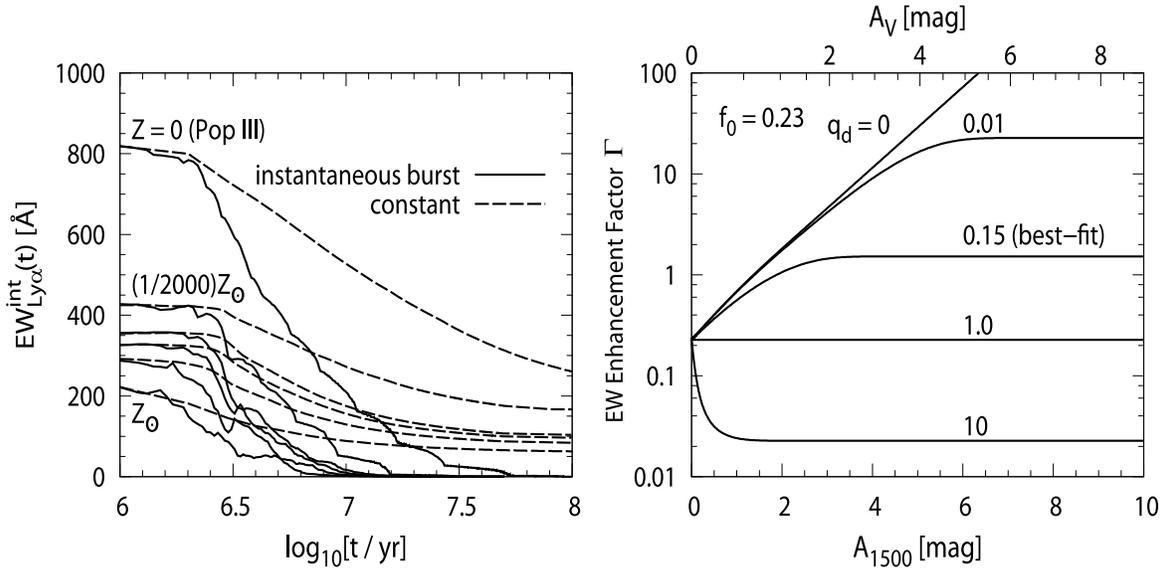} 
 \caption{ 
   \textit{Left}: the time evolution of $\ewint$ (``intrinsic'' $\lya$
   EW determined by stellar spectra and the case B recombination with
   the assumption all of LyC photons are absorbed and reprocessed into
   Ly$\alpha$ photons in \ion{H}{2} regions) assuming the S03
   population synthesis model with the Salpeter IMF.  The solid and
   dashed curves represent the $\ewint$ evolution for instantaneous
   starburst and constant star formation, respectively. Several curves
   are shown corresponding to different stellar metallicities:
   $Z/Z_\odot = 0$, $1/2000$, $1/50$, $1/20$, $1/5$, and $1$, from top
   to bottom.
   \textit{Right}: the EW enhancement factor $\Gamma$ as a function of
   the extinction magnitude at 1500~{\AA}, $A_{1500}$, or at V-band,
   $A_V$.  We plot $\Gamma$ for several values of $q_d$ (including the
   value of $q_d=0.15$ used in our LAE model) in the range of
   $q_d=0-10$ as indicated in the figure, while $f_0$ is fixed to
   $0.23$ (the value used in the model).
 }
 \label{fig-EW-Gamma}
\end{figure}

  Next we calculate $\lya$ EW of each galaxy in the model described
  above.  In addition to the \textit{observable} EW (i.e., dust
  extinction incorporated into both luminosities of $\lya$ line and UV
  continuum) denoted as $\ewlya$, it is convenient to define
  \textit{intrinsic} $\ewint$ that is simply calculated using the S03
  stellar spectra and the Salpeter IMF, assuming the case B
  recombination and 100\% escape fractions both for $\lya$ and UV
  continuum photons.  Note that all EW values quoted in this paper are
  in rest frame.  We need to calculate UV continuum luminosity at
  $\lambda = \lambda_{\rm Ly\alpha} = 1216$~{\AA} to estimate EW.  We
  calculate it from the UV luminosity at 1500~{\AA} and assuming a
  spectral index of $\beta = -2$, where $f_\lambda \propto
  \lambda^\beta$ (i.e., a flat spectrum in $f_\nu$), for the stellar
  spectra without extinction by dust.  This is a typical UV spectrum
  of young stellar population (Bruzual \& Charlot 1993; Leitherer et
  al. 1999), and it is often assumed in studies of high-redshift
  galaxies including LAEs (e.g., Ouchi et al. 2008).  The time
  evolutions of $\ewint$ in the two cases of instantaneous starburst
  population and constant star formation are plotted in
  Fig.~\ref{fig-EW-Gamma} (left panel), for some values of
  metallicities.  The maximum values of $\ewint$ are obtained at an
  age of $\sim 1$~Myr after the onset of star formation, with the
  values of $\sim 240$, 420, and 820~{\AA} for metallicities of
  $Z/Z_\odot = 1$, $1/2000$, and 0 (Pop III), respectively.
  
  However, because of the $\lya$ photon escape fraction and extinction
  of continuum photons, $\ewlya$ can be different from $\ewint$.  From
  the modeling above, the ratio is given as:
  \begin{equation}
   \Gamma \equiv \frac{\ewlya}{\ewint} = \frac{f_0}{q_d} 
    \frac{ 1 - \exp[ - q_d \taucont(\lambda_{\rm Ly\alpha}) ] }
    { 1 - \exp[ - \taucont(\lambda_{\rm Ly\alpha}) ] }
    \ .
    \label{eq-gamma1}
  \end{equation}
  (Note that $\Gamma = f_0^{\rm wind}
  \taucont(\lambda_\mathrm{Ly\alpha}) /\left\{ 1 -
  \exp[-\taucont(\lambda_\mathrm{Ly\alpha})]\right\}$ for starburst
  galaxies in the outflow phase.)  When $q_d < 1$, continuum photons
  are more significantly absorbed by dust than $\lya$ photons, and
  $\ewlya$ can be larger than $\ewint$ ($\Gamma > 1$).  In our model,
  although $q_d < 1$, the EW enhancement effect is reduced by the
  dust-independent factor of $f_0$.  We plot the EW enhancement factor
  $\Gamma$ against the reddening parameter $A_{1500}$ in
  Fig.~\ref{fig-EW-Gamma} (right panel) for several values of $q_d$.
  When $q_d < 1$, $\Gamma$ increases with $A_{1500}$, and then reaches
  an asymptotic value of $f_0/q_d$.  In the case of our model ($f_0 =
  0.23$ and $q_d = 0.15$), the asymptotic value becomes $\Gamma =
  1.53$, i.e., only a modest enhancement of EW.  (It is interesting
  that a different theoretical model of Dayal et al. (2008)
  independently predicted a similar value of $\sim 1.6$.)  Therefore,
  in our model, we need intrinsically large EW LAEs having $\ewint
  \geq 160$~{\AA} to explain LAEs having $\ewlya \geq 240$~\AA.
  However, $\ewlya$ is strongly suppressed by a factor of $\Gamma \sim
  f_0 = 0.23$ (or $f_0^{\rm wind} = 0.36$) in the case of
  $A_{1500}\sim 0$, and hence the dust extinction effect should also
  have an important role to achieve large $\lya$ EWs.
  
  Our result indicating the importance of dust extinction for large EW
  LAEs is consistent with recent observational results that some LAEs
  indeed seem to have large EWs by the effect of clumpy dust
  distribution (Finkelstein et al. 2008, 2009a, 2009b).  It should be
  noted that the parameter $q_d$ in this work is slightly different
  from an observational estimate of the geometry parameter $q_{\rm
  obs}$ introduced by Finkelstein et al. (2008).  The
  dust-independent factor $f_0$ is not taken into account in $q_{\rm
  obs}$, and it is defined as $q_{\rm obs} = \tau_d^{\rm Ly\alpha} /
  \taucont(\lambda_{\rm Ly\alpha})$ assuming screen geometry
  (attenuation $\propto e^{-\tau}$) in the SED fits.  Our definition
  of $q_d$ becomes equivalent to $q_{\rm obs}$ only when we set $f_0 =
  1$ and change the geometry of dust distribution from slab to screen.

%%%%%%%%%%%%%%%%%%%%%%%%%%%%%%%%%%%%%%%%%%%%%%%%%%%%%%%%%%%%%%%%%%%%%
 \section{COMPARISONS WITH OBSERVATIONS}
 \label{section:comparison} 
 
  \subsection{Definition of LAEs and Selection Criteria}
  \label{subsec:stat}
  
  In observations of LAEs, there are mainly two criteria to select LAE
  candidates from photometric samples: the limiting magnitude of
  narrow-band filter that catches redshifted $\lya$ lines and the
  color between narrow- and broad-band filters.  These roughly
  correspond to the criteria of $\lya$ luminosity ($\lobs \geq \lth$)
  and $\lya$ EW ($\ewlya \geq \ewth$), respectively.  Different
  criteria are applied for different observations, and it is important
  to compare the model with observations under appropriate treatments
  of these criteria.  In this work, we always select model LAEs by the
  same threshold values of $\lth$ and $\ewth$ as those adopted in each
  observation.  The threshold values used in various LAE observations
  to be compared with our model in this paper are compiled in
  Table~\ref{tab-comp}.  
  \placetable{tab-comp}

  \subsection{Ly$\alpha$ Luminosity Functions}
  \label{subsec:lyalf}

\begin{figure}
 \epsscale{1.}
 \plotone{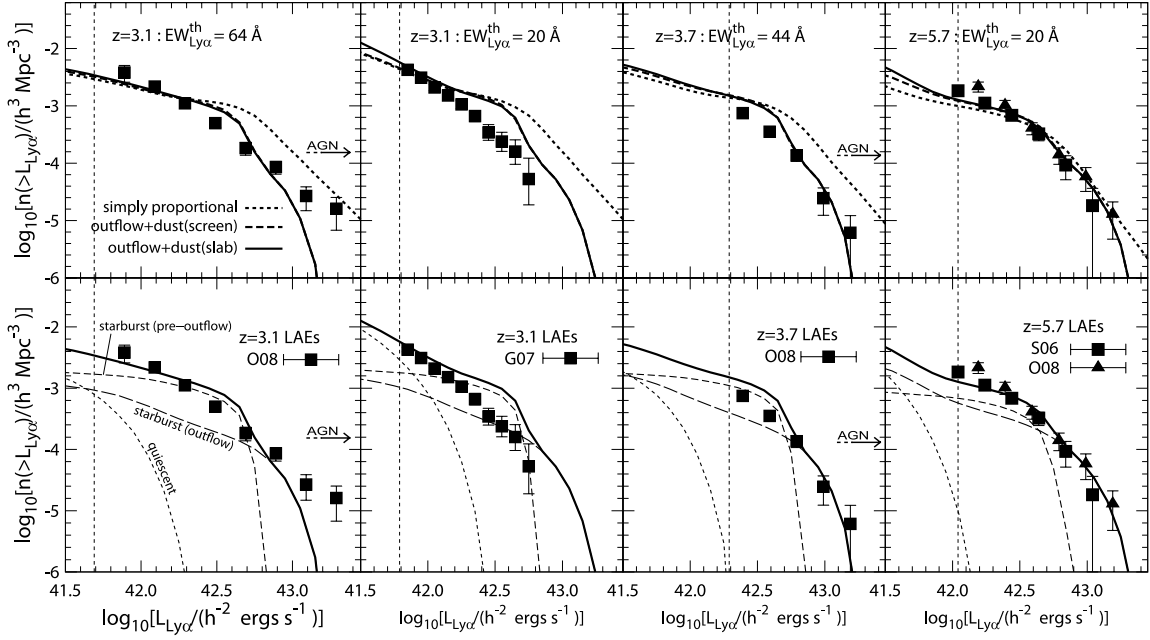}
 \caption{ 
   Cumulative LAE $\lya$ LFs at $z\sim 3$--6.  The model predictions
   are represented by the curves, while the observational data are
   plotted by the filled symbols with error-bars.  At each panel, the
   model LAEs are selected with the same threshold value of $\ewth$ as
   that adopted in each observation, which is indicated in each panel.
   The vertical dashed line in each panel is the threshold value of
   $\lth$ of each observation.  
   \textit{Top}: the predictions from three different $\fesclya$ model
   are shown; the simply proportional, the outflow$+$dust (screen),
   and the outflow$+$dust (slab) models are represented by the dotted,
   dashed, and solid curves, respectively.
   \textit{Bottom}: the contributions from quiescent, pre-outflow
   phase starburst, and outflow phase starburst are plotted separately
   by the dotted, short-dashed, and long-dashed curves, respectively,
   in the outflow$+$dust (slab) model.  The references for data points
   are Shimasaku et al. (2006), Gronwall et al. (2007), and Ouchi et
   al. (2008).  The arrows are taken from Ouchi et al. (2008) which
   represent the $\lya$ luminosity range where $\lya$ LFs are
   dominated by LAEs with AGN activities.
 }
 \label{fig-LyALF-LAE}
\end{figure}

  In Fig.~\ref{fig-LyALF-LAE}, we show our model predictions for the
  $\lya$ LF of LAEs, in comparison with observations at $z\sim $3--6.
  This comparison was already done by KTN07 in detail and with more
  observed data, but here we show only the observed data that also
  have UV luminosity and EW data used in this paper.  As mentioned in
  the previous section, the thresholds for $\lobs$ and $\ewlya$ are
  important in this kind of comparison, and here we correctly match
  these conditions between the model and the data.  Two different
  panels are shown for the same redshift of $z = 3.1$, corresponding
  to different observational data by different authors using their own
  LAE selection criteria.  On the other hand, the LAE selection
  criteria of the $z=5.7$ data by Shimasaku et al. (2006) and Ouchi et
  al. (2008) are similar, and we plot only one model in comparison
  with them.

  The overall levels and characteristic break luminosities of $\lya$
  LFs predicted by the outflow$+$dust (slab) model (the model used in
  this work) are in reasonable agreement with the observations.  The
  agreement with the S06 data at $z=5.7$ is rather trivial, because we
  determined the three LAE model parameters by fitting to these data.
  The bright-end cut-off in the model LF for $z=3.1$ selected with the
  criteria of Ouchi et al. (2008) is too sharp compared with the data.
  However, according to Ouchi et al. (2008), there is a significant
  contamination by AGNs at $z=3.1$ and $z=3.7$ in the brightest
  luminosity range of $\lobs \gtrsim 10^{43}~ h^{-2}~ \mathrm{ergs~
  s^{-1}}$, while no significant AGN contamination was found at $z =
  5.7$.  This might be a possible reason for this discrepancy.
  Quiescent galaxies contribute only to relatively low $\lobs$ LAEs,
  and their contribution becomes smaller with increasing redshift.
  Therefore, their contribution is negligible if the limiting
  magnitude of narrow band is shallow (i.e., $\lth \gtrsim 10^{42}~%
  h^{-2}~ \mathrm{ergs~ s^{-1}}$).

  The alternative models (the simply-proportional model and the
  outflow$+$dust (screen) model) tested in KTN07 are also shown. The
  difference between the slab and screen dust is negligible, while the
  simply-proportional model overproduces the bright-end of $\lya$ LFs
  especially at lower-$z$.

  \subsection{UV Continuum Luminosity Functions}
  \label{subsec:uvlf}
  
\begin{figure}
 \epsscale{1.}
 \plotone{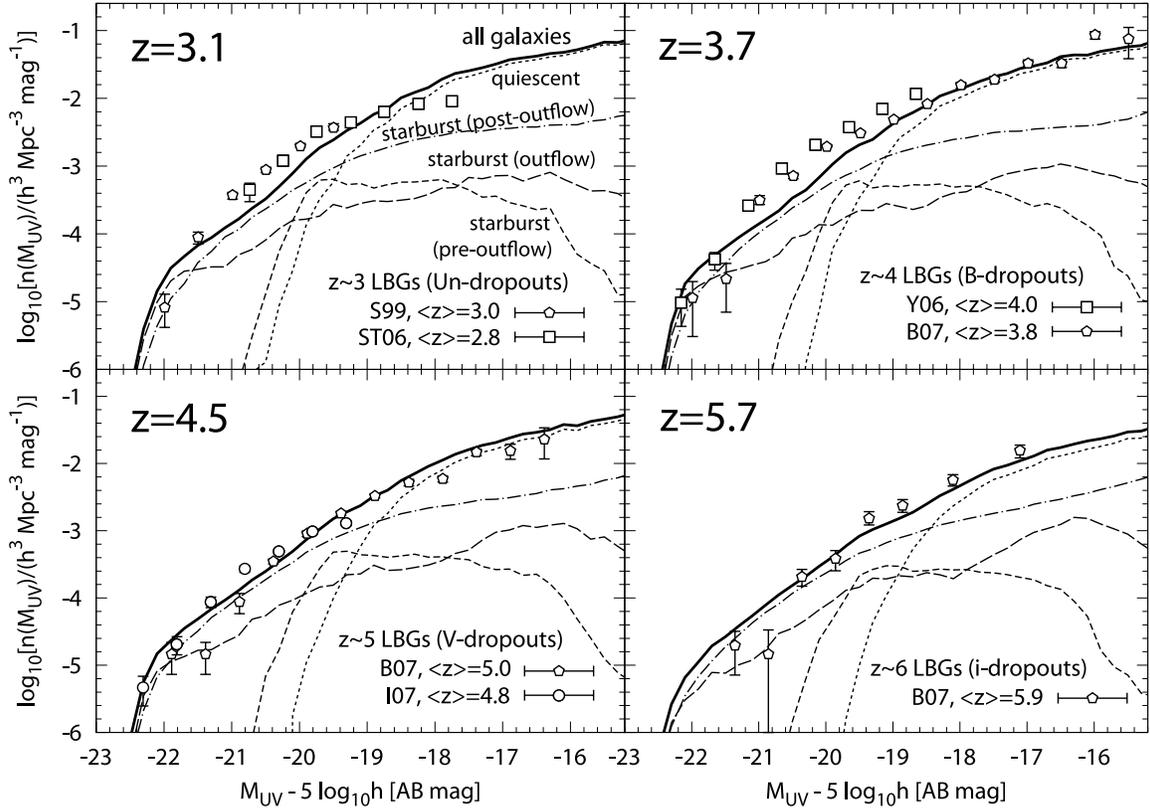}
 \caption{ 
   Rest-frame UV ($\lambda =1500$~\AA) LFs of all (i.e., including
   non-LAEs) galaxies at $z \sim 3$--6.  The curves are predictions by
   our model, while the open symbols with error-bars are the observed
   data of the LBGs at similar redshifts.  The contributions from the
   quiescent galaxies and the starbursts in the pre-outflow, outflow,
   and post-outflow phases are shown separately by the dotted,
   short-dashed, long-dashed, and dash-dotted curves, respectively.
   Note that post-outflow galaxies are assumed to produce no $\lya$
   photon and hence do not contribute to the model LAE population, but
   contribute to LBGs by UV continuum (see text).  The references for
   the data points of LBGs are Steidel et al. (1999), Sawicki \&
   Thompson (2006), Yoshida et al. (2006), Bouwens et al. (2007), and
   Iwata et al. (2007).
 }
 \label{fig-UVLF-LBG}
\end{figure}

  Before showing UV continuum LFs of LAEs, we examine whether our
  model correctly reproduces the UV LF of LBGs, which are a more
  abundant and well studied population of high redshift galaxies.
  Because of their simpler selection criteria based on the Lyman break
  absorption feature by neutral hydrogen in the IGM and ISM (e.g.,
  Yoshii \& Peterson 1994; Madau 1995), UV LF of LBGs can practically
  be considered as that of all galaxies at the redshift.  We show the
  rest-frame UV LFs of all model galaxies (i.e., no selection) at $z
  \sim $ 3--6 in Figure~\ref{fig-UVLF-LBG}.  Our model are in
  reasonable agreement with the observations.

\begin{figure}
 \epsscale{1.}
 \plotone{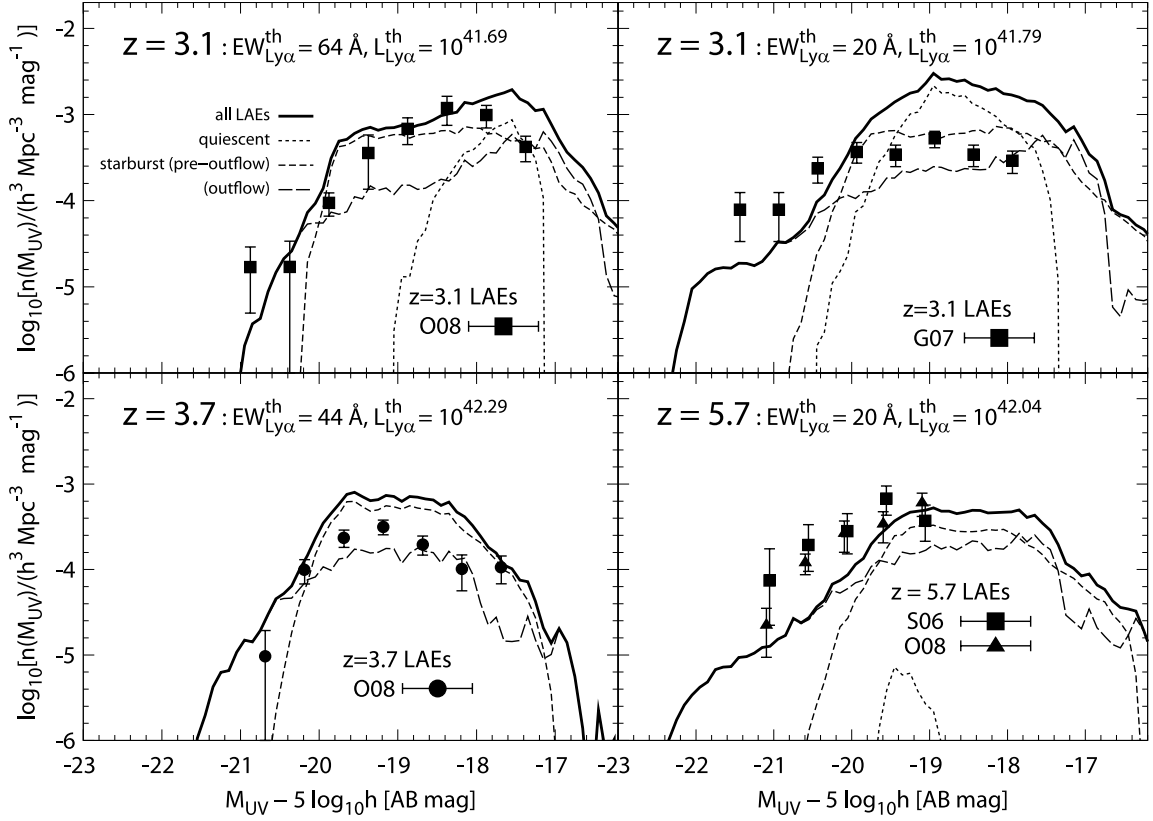}
 \caption{
   The same as Fig.~\ref{fig-UVLF-LBG} but for LAE UV LFs.  The curves
   are the model predictions, while the solid symbols with error-bars
   are the observed data of the LAEs.  The model LAEs have been
   selected with the same values of $\lth$ and $\ewth$ with those
   adopted in each observation shown in each panel (the unit of $\lth$
   is $h^{-2}\ \mathrm{ergs\ s^{-1}}$).  Both of the top panels are
   the UV LFs of the LAEs at $z=3.1$, but they are selected with
   different criteria.  The line markings are the same as
   Fig.~\ref{fig-UVLF-LBG}.  The references for the data points of
   LAEs are Shimasaku et al. (2006), Gronwall et al. (2007), and Ouchi
   et al. (2008).
 }
 \label{fig-UVLF-LAE}
\end{figure}

  We now turn to the UV LFs of LAEs.  Figure~\ref{fig-UVLF-LAE}
  presents comparisons of LAE UV LFs at $z\lesssim 6$ between model
  predictions and observational results.  
  \placefigure{fig-UVLF-LAE}
  For the LAEs at $z=3.1$, we plot two different UV LFs for the model
  LAEs as done in Fig.~\ref{fig-LyALF-LAE}.  We find that, although
  the $z=3.1$ data points by Gronwall et al. (2007) and the $z=3.7$
  points by Ouchi et al. (2008) show quantitative discrepancies (up to
  a factor of a few) at low UV luminosity range ($M_{\rm UV} -5\log{h}
  \lesssim -19$~mag) in comparison with the model predictions, the
  overall profiles are roughly reproduced.

  \subsection{$\lya$ EW Distributions and the Origin of Large EW LAEs}
  \label{subsec:EW}
  
\begin{figure}
 \epsscale{1.}
 \plotone{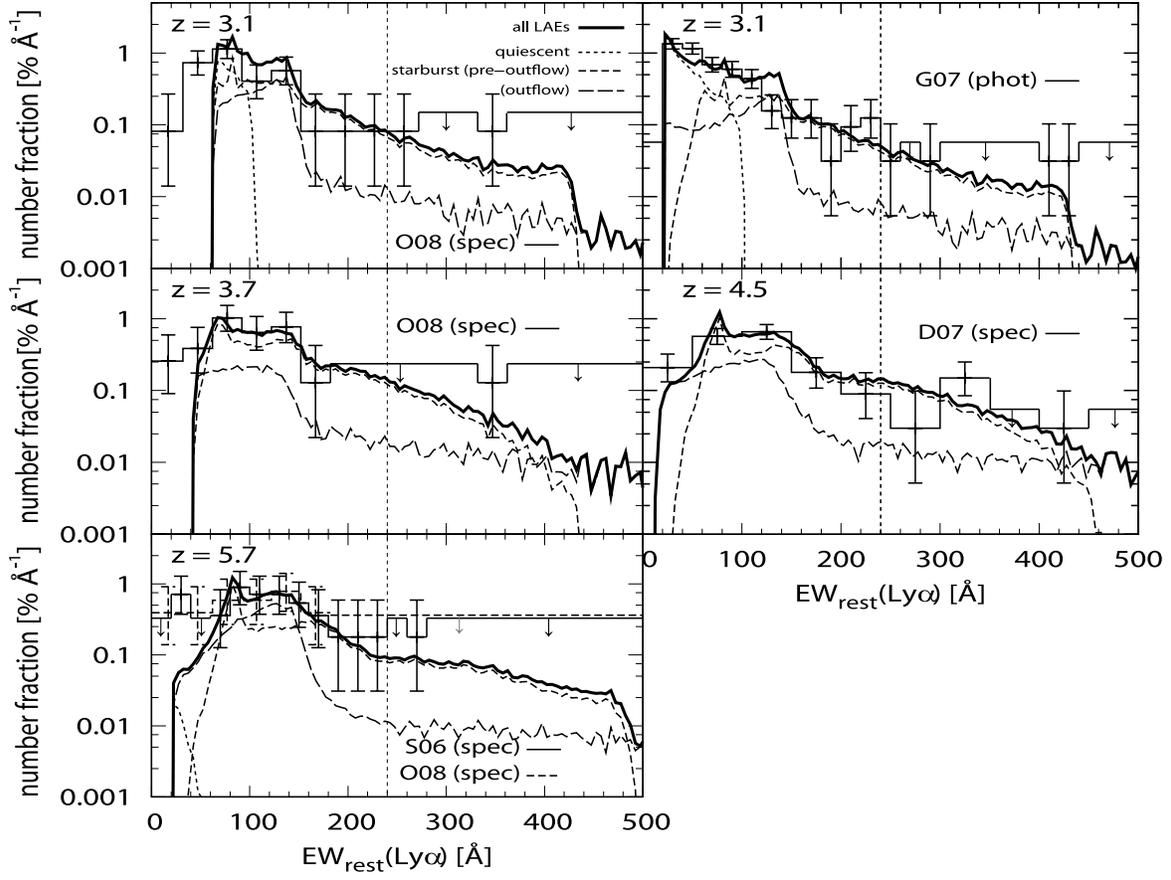}
 \caption{ 
   Rest-frame EW distributions for LAEs at $z\sim 3$--6.  The curves
   are predictions by our model, while the histograms are the
   observations.  The observed data based on spectroscopically
   confirmed LAEs are labeled as ``spec'', while those based on
   photometric samples as ``phot''.  The error bars and upper limits
   are the $1\sigma$ Poissonian statistics for small sample number
   (Gehrels 1986).  The line markings of the model curves are the same
   as Fig.~\ref{fig-UVLF-LBG}.  The vertical dashed line indicates the
   maximum $\ewint$ value (240~\AA) powered by star-formation activity
   with the Salpeter IMF and the solar metallicity (Charlot \& Fall
   1993; Schaerer 2003).  The references of the data are given in
   Fig.~\ref{fig-UVLF-LAE}, except for Dawson et al. (2007).
 }
 \label{fig-EW}
\end{figure}

  The observed EW distributions by a variety of authors at $z \sim
  3$--6 are shown in Figure~\ref{fig-EW}.  
  \placefigure{fig-EW}
  The statistical 1$\sigma$ errors and upper limits are calculated by
  the small number Poisson statistics tabulated by Gehrels (1986).
  Our model predictions are also presented for comparison.  A small
  number of observed galaxies have $\ewlya$ smaller than the threshold
  $\ewth$, while the model predicts exactly $\ewlya \geq \ewth$.  This
  is because the criteria to select LAE candidates from photometric
  samples in actual observations are not exactly the same as those in
  the model based only on $\lth$ and $\ewth$.  However, the difference
  is not significant and does not affect our main conclusions in this
  paper.
  
  The model predictions are in good agreement with the observed
  distributions at various redshifts.  It should be noted that our
  model naturally reproduces the observed EW distributions well beyond
  $\ewlya \gtrsim 240$~\AA, even though our model does not include Pop
  III stars and it assumes the ordinary Salpeter IMF.  This indicates
  that Pop III stars or an extremely top-heavy IMF are not inevitably
  required to interpret the existence of large EW LAEs.
  
\begin{figure}
 \epsscale{1}
 \plotone{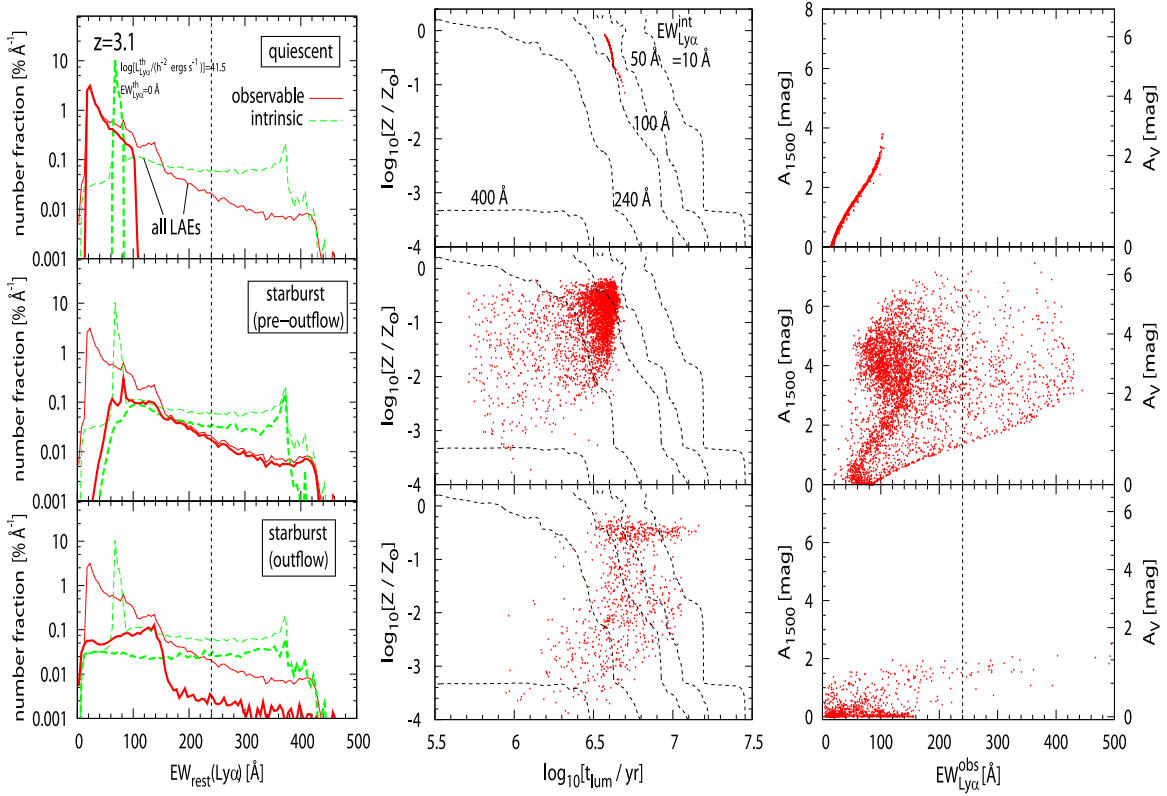}
 \caption{ 
   The left, middle, and right panels show EW distributions (intrinsic
   and observable), luminosity-weighted age ($t_{\rm lum}$) versus
   metallicity, and observable EW versus extinction, respectively, of
   the model LAEs at $z=3.1$.  The quantity $t_{\rm lum}$ represents a
   typical age of stellar population that is responsible for $\lya$ EW
   (see text for exact definition).  The contributions from quiescent,
   pre-outflow phase starburst, and outflow phase starburst are shown
   separately from top to bottom panels.  In the left panels, total EW
   distributions including all of the three populations (quiescent,
   pre-outflow, and outflow) are also shown by thin curves.  The small
   dots in the middle and right panels are model galaxies.  In the
   middle column, the contours of $\ewint$ of instantaneous starburst
   population (shown in Fig.~\ref{fig-EW-Gamma}) are shown.  The model
   LAEs are selected with $\lth =10^{41.5}\ h^{-2}\ \mathrm{ergs\
   s^{-1}}$, but with no selection about $\ewlya$.
 }
 \label{fig-ch}
\end{figure}

  \placefigure{fig-ch} 
  To understand the reason why our model could reproduce the large EW
  LAEs, we plot the EW (intrinsic as well as observable, as defined in
  \S~\ref{subsec:model_EW}) distributions in the left column of
  Figure~\ref{fig-ch}, for the quiescent, pre-outflow starburst, and
  outflow starburst populations.  Here, we show the case of $z = 3.1$
  as an example, with no selection about EW (i.e., $\ewth = 0$~\AA)
  but $\lth = 10^{41.5}~ \unitlum$.  The distributions of intrinsic EW
  can be understood by the distributions of metallicity and
  characteristic stellar age of model galaxies, which are shown in the
  middle column of the same figure.  Here, the ``characteristic age''
  is defined as that of the stellar populations mainly contributing to
  the values of EW, so that $\ewint$ of the model galaxies should be
  close to those of instantaneous starburst having the same
  metallicity and age.  For comparison, the contours of $\ewint$ of
  instantaneous starbursts are depicted in the same panels, giving an
  approximate $\ewint$ values of the model galaxies in this plane.
  However, the definition of the characteristic age is rather
  complicated, and we explain below.
  
  EW is determined by ionizing luminosity ($\lambda < $ 912~{\AA}
  and proportional to the intrinsic $\lya$ luminosity) and continuum
  luminosity around 1216~\AA, and hence two characteristic time scales
  can be defined: the mean stellar age weighted by ionizing
  luminosity, $t_{\rm lum}^{\rm ion}$, and the mean stellar age
  weighted by UV continuum luminosity, $t_{\rm lum}^{\rm UV}$.  We
  found that $t_{\rm lum}^{\rm UV}$ becomes significantly greater than
  $t_{\rm lum}^{\rm ion}$ especially at ages of $t_{\rm lum}^{\rm
  ion}\sim 10^{6.5}$~yr, and $\ewint$ values of model galaxies are
  considerably different from those of instantaneous starbursts at
  either age of $t_{\rm lum}^{\rm ion}$ or $t_{\rm lum}^{\rm UV}$.
  Therefore, we take the harmonic mean\footnote{We found that the EW
  of instantaneous starburst population at the harmonic mean age is
  closer to the model EWs, than the arithmetic or geometric means.}
  of the two ages to calculate the characteristic age $t_{\rm lum}$,
  which is the quantity plotted in the middle column of
  Fig.~\ref{fig-ch}.  Then these plots show the effects of age and
  metallicities in the resultant values of $\ewint$ in each model
  galaxy.
  
  The intrinsic EW distribution of the quiescently star-forming
  galaxies is narrowly peaked at $\ewint \approx 70$~\AA.  This is
  because these galaxies are forming stars approximately constantly on
  time scales longer than the lifetimes of massive stars contributing
  to the ionizing luminosity or continuum luminosity around 1216~\AA,
  and thus EW reaches an equilibrium value determined by stellar
  metallicity and IMF (Charlot \& Fall 1993; Schaerer 2003).  Although
  metallicities of the model galaxies have some scatter in a range of
  $Z/Z_\odot \sim 10^{-1.5}$--$1$, equilibrium EWs are narrowly
  concentrated to $\ewint \approx 70$~{\AA} in this metallicity
  range. (See Fig.~\ref{fig-EW-Gamma} and the middle column of
  Fig.~\ref{fig-ch}.)  In spite of the narrow width of $\ewint$
  distribution, $\ewlya$ are distributed more widely because model
  galaxies have a variety of extinction amount, as shown in the right
  column of Fig.~\ref{fig-ch}.  Since $\ewint$ is limited to be
  $\lesssim 100$~\AA, the quiescent galaxy population cannot explain
  galaxies having $\ewlya \gtrsim 240$~\AA.
  
  On the other hand, the distribution of intrinsic EW extends to a
  very large value of $\sim 400$~{\AA} in the case of pre-outflow
  starbursts, because of small stellar ages and low metallicities (see
  the middle row of Fig.~\ref{fig-ch}).  The observed EW distribution
  also extends to $\sim 400$~\AA, and it is this LAE population that
  contributes to the large EW LAEs of $\gtrsim 240$~\AA.  However, it
  should be noted that simply large $\ewint$ is not sufficient to have
  large $\ewlya$.  A significant amount of dust (i.e., $A_{1500}
  \gtrsim 2$~mag) is also required to compensate the EW reduction by
  the dust-independent factor $f_0$.  The large $\ewlya$ found for
  this population is due to the combined effects of stellar population
  (small age and low metallicity) and clumpy dust.

  The bottom panels of Fig.~\ref{fig-ch} shows the same but for
  outflow starbursts.  As shown in the bottom-left panel of
  Fig.~\ref{fig-ch}, small fraction of the galaxies also contributes
  to the large EW LAEs at $z=3.1$.  They are found to be so young that
  their $\ewint$ are large and that non-negligible amount of dust
  remains in their ISM, which can enhance their $\ewint$ to $\ewlya
  \gtrsim 240$~\AA.  Hence, the importance of the combined effects of
  stellar population and clumpy dust is revealed again.

  \subsection{Correlation between Ly$\alpha$ EW and Extinction}
  
  The above results suggest that extinction by clumpy dust is playing
  an important role to produce LAEs having large EWs.  Then we expect
  some correlations between observable EWs and reddening.  Therefore
  we plot $\ewlya$ vs. $A_{1500}$ (or $A_V$) in the right panels of
  Fig.~\ref{fig-ch}.  The quiescent galaxies populate the lower-left
  corner (small EWs and $A_{1500}$), and a clear correlation between
  $\ewlya$ and $A_{1500}$ is found, because $\ewint$ has a narrow
  distribution and the difference of $\ewlya$ is due to the difference
  of extinction.  The pre-outflow and outflow starburst galaxies have
  a large scatter in this plot, because of the wide distribution of
  $\ewint$.  However, there is a forbidden region in the lower-right
  corner, meaning that large EW LAEs ($\ewlya \gtrsim 200$~\AA) must
  have non-negligible extinction ($A_{1500}\gtrsim 1$~mag).  Then we
  expect some correlation or trend between $\ewlya$ and $A_{1500}$,
  depending on the relative proportions of these three populations.
  
\begin{figure}
 \epsscale{1}
 \plotone{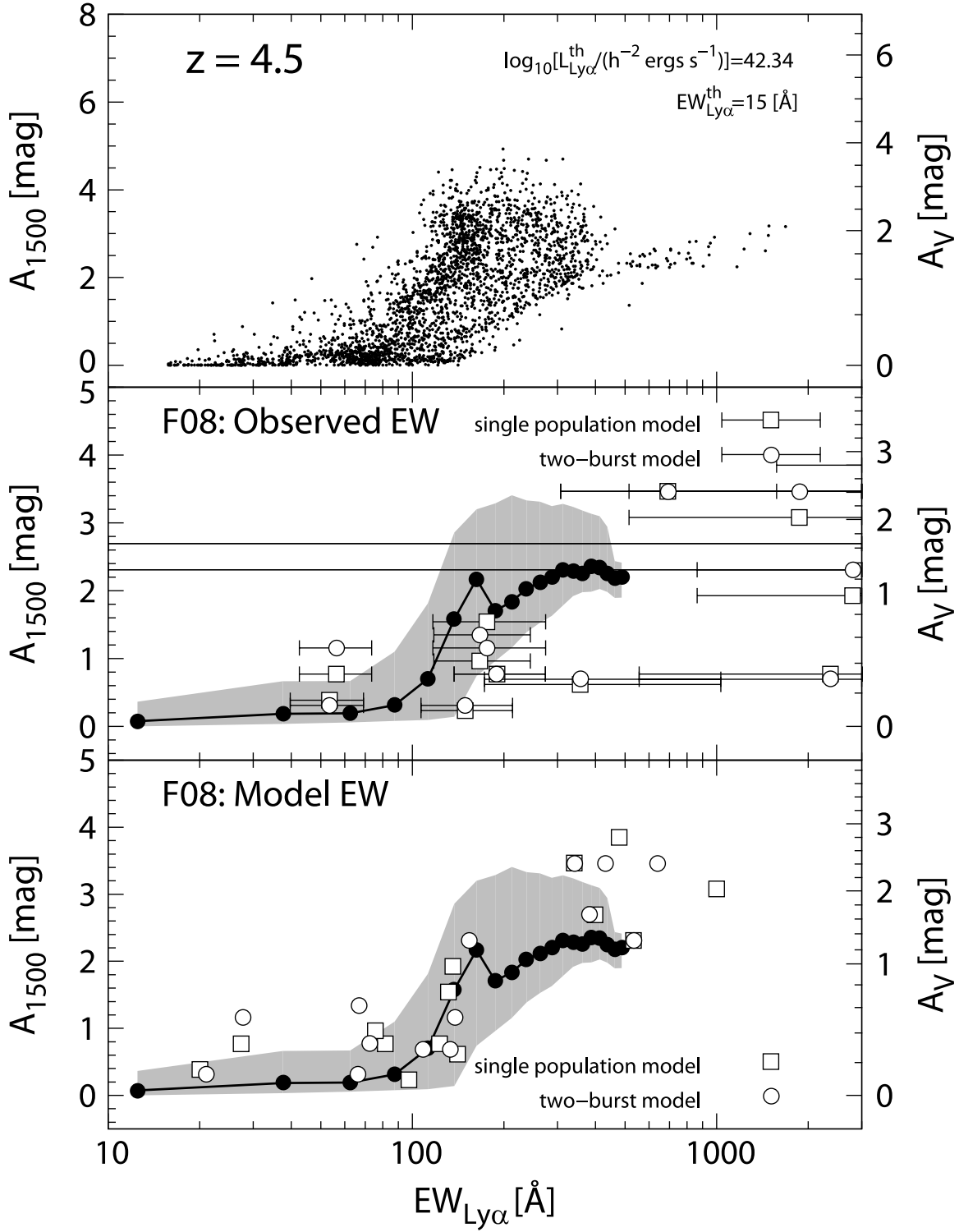}
 \caption{
   $A_{1500}$ (\textit{left ordinate}) or $A_V$ (\textit{right
   ordinate}) versus $\ewlya$.  The distribution of model LAEs at
   $z=4.5$ are represented by small dots in the top panel.  The
   selection criteria of model LAEs are indicated in the top panel,
   which are identical to those adopted in Finkelstein et al. (2009a).
   In the middle and bottom panels, the median of the model LAE
   distribution is shown by the filled circles connected by solid
   lines, while the $10-90$ percentiles are represented by the gray
   regions.  The open symbols with error-bars are observational data
   of 14 LAEs at $z\sim 4.5$ given by Finkelstein et al. (2009a).  The
   data of $A_{1500}$ were derived by their SED fittings and its
   typical error is $\sim 0.8$~mag (S. Finkelstein, private
   communication).  The $\ewlya$ data in the middle panel were derived
   by direct observational measurements, while those in the bottom
   panel were estimated by their SED fittings.  The open squares and
   circles represent the best-fit values using the single population
   model and two-burst model, respectively, in their SED fittings.
 }
 \label{fig-EW-A1200-F08}
\end{figure}

  In fact, recent observations by Finkelstein et al. (2008, 2009a)
  have indicated that the extinction by clumpy dust has an important
  effect to produce large EW LAEs at $z=4.5$, by comparing EWs and
  $A_{\rm 1200}$ estimated from the SED fitting.  The data set of
  Finkelstein et al. (2009a) can directly be compared with our model
  prediction in the $\ewlya$-$A_{1500}$ plane, which is given in
  Fig.~\ref{fig-EW-A1200-F08}. (The values of $A_{1200}$ in
  Finkelstein et al. have been converted to $A_{1500}$ by the
  extinction law assumed in this work).  Direct observational
  measurements of $\ewlya$ by Finkelstein et al. (2009a) suffer from
  large uncertainties for galaxies with faint UV flux (or, large EWs),
  and hence we also plot model-EWs that are estimated by Finkelstein
  et al. (2009a) using the UV luminosities calculated by the SED
  fitting rather than observationally measured UV luminosities.
  Because of the degeneracy between dust extinction and stellar age,
  the estimates of $A_{1500}$ typically have a $1\sigma$ error of
  $\lesssim 0.8$~mag (S. Finkelstein, private communication).  When
  the three model populations are mixed with the LAE selection
  conditions matched to theirs, we find that our model distribution is
  in good agreement with that of the observed data.  Here, the
  contribution of the quiescent galaxies is negligible because of
  their relatively large threshold $\lya$ line luminosity, $\lth =
  10^{42.34}\ h^{-2}\ \mathrm{ergs\ s^{-1}}$.  
  
  The positive correlation among $A_{1500}$ and $\ewlya$ seems to be
  inconsistent with the results for the LBGs at $z\sim 3$ (Shapley et
  al. 2003) and those at $z\sim 3.5-6$ (Pentericci et al. 2009)
  because they reported that more dust extincted LBGs have lower EW on
  average.  However, it should be mentioned here that their results do
  not contradict our prediction.  This is because their samples are
  limited at relatively luminous range of UV continuum, $M_{\rm UV} -
  5\log{h} \lesssim -19$~mag.  In our model, such UV luminous galaxies
  have small EW ($\mathrm{EW} \lesssim 150$~\AA) and less dust
  extinction ($A_{1500} \lesssim 2$~mag).  The positive correlation
  predicted by our model is not clearly seen in such a small EW and
  $A_{1500}$ range as presented in Fig.~\ref{fig-EW-A1200-F08}.
  Observations of the galaxies with $M_{\rm UV} - 5\log{h} \gtrsim
  -19$~mag is required to test the validity of our model prediction.

  This result gives a further support to the idea of EW enhancement by
  clumpy dust, which have been inferred from the observed
  EW-$A_{1500}$ correlation, and independently, from the LAE $\lya$ LF
  modeling in our previous work.  Although the statistics is still
  limited, comparisons in this $\ewlya$-$A_{1500}$ plane with more
  data obtained in the future will provide an interesting test of our
  model for LAEs.

  \subsection{$M_{\rm UV}$-EW Plot and the Ando Effect}
  \label{subsec:AndoPlane}
  
\begin{figure}
 \epsscale{1.}
 \plotone{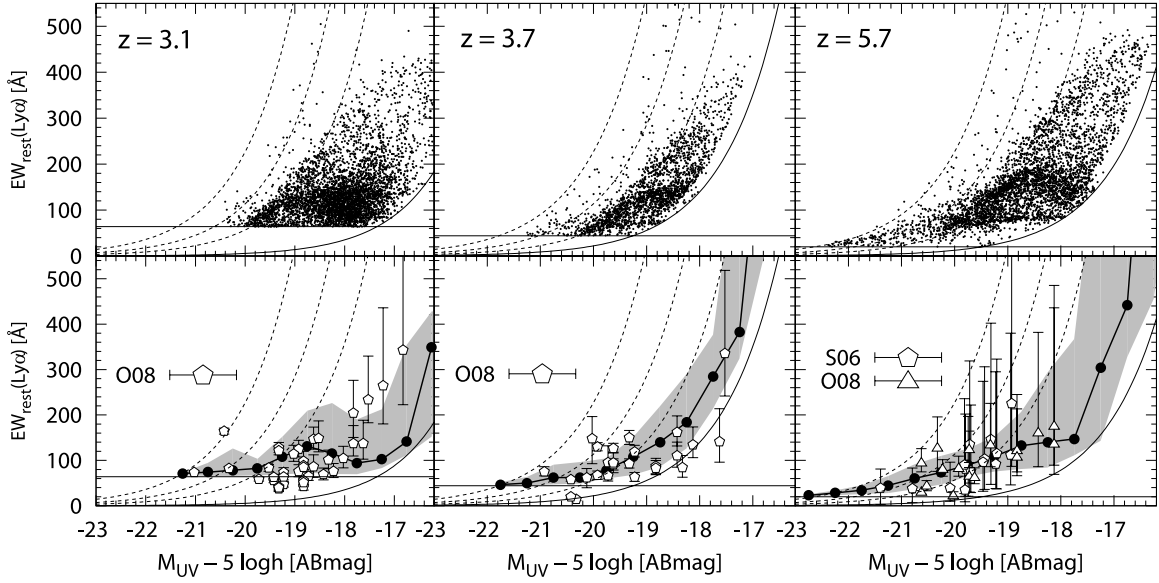}
 \caption{ 
   Distributions of LAEs in the $M_{\rm UV}$-$\ewlya$ plane at $z\sim
   $3--6.  In the upper panels, model LAEs are plotted by small dots.
   In the lower panels, the open symbols are corresponding to each of
   the observed galaxies. The filled circles connected by solid lines
   are the mean of the model galaxies, and the gray region represents
   the $10-90$ percentiles of the model galaxies.  See
   Fig.~\ref{fig-UVLF-LAE} for the references.  The solid curve and
   horizontal line in each panel indicate $\lth$ and $\ewth$,
   respectively. Note that $M_{\rm UV}$ is in the rest-frame 1500~\AA,
   and we must convert it in the rest-frame $\lya$ wavelength to
   depict the curve of $\lth$. We assumed a SED with $\beta = -2$ as
   an approximation here.  The dotted curves show the contours of
   $\lya$ line luminosity corresponding to $2\times 10^{43}$,
   $10^{43}$, and $5\times 10^{42}\ \unitlum$ from top left to bottom
   right, assuming $\beta = -2$.
 }
 \label{fig-Ando}
\end{figure}

\begin{figure}
 \epsscale{1.}
 \plotone{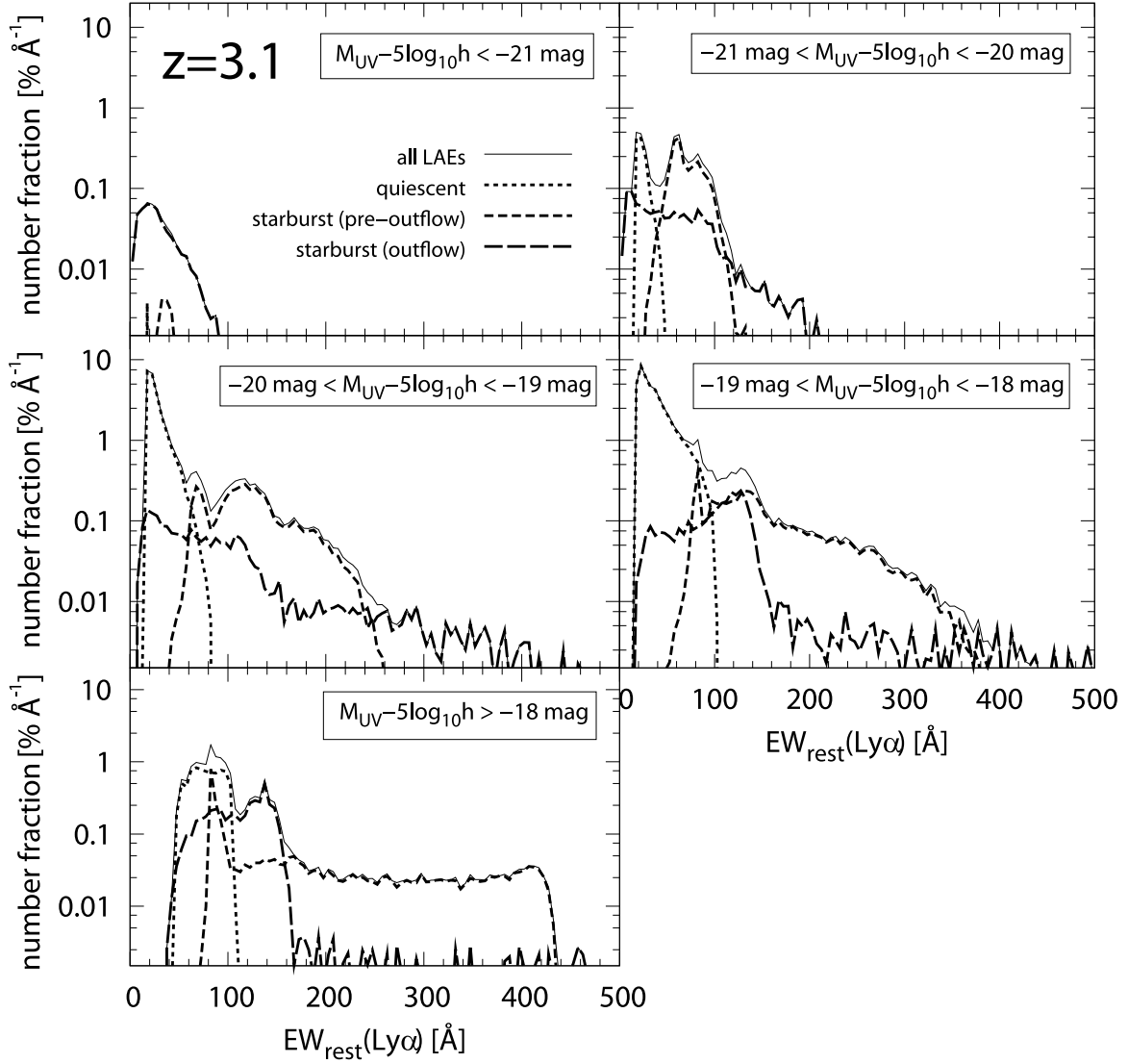}
 \caption{ 
   EW distribution for various $M_{\rm UV}$ intervals at $z=3.1$.
   Note that $M_{\rm UV}$ is the observable (i.e., extinction
   uncorrected) one.  The selection criteria of model LAEs are $\lth
   =10^{41.5}\ h^{-2}\ \mathrm{ergs\ s^{-1}}$ and $\ewth = 0$~{\AA}.
   The line markings are indicated in the upper-left panel (the same
   as Fig.~\ref{fig-UVLF-LBG}).
 }
 \label{fig-EWdist-L_UV}
\end{figure}

  \placefigure{fig-Ando} 
  In Figure~\ref{fig-Ando}, the comparisons between the model
  predictions and the observational data in the $M_{\rm UV}$-$\ewlya$
  plane are presented.  Note that $M_\mathrm{UV}$ is observable
  (i.e. dust extinction uncorrected) absolute magnitude of UV
  continuum.  The model LAEs distribute similarly to the observed LAEs
  in this plane.  Particularly, the trend of the Ando effect (smaller
  EW for brighter $M_{\rm UV}$) is reproduced well by our model.  To
  see the reason why our model reproduces the Ando effect correctly,
  we show the EW distributions for several different $M_{\rm UV}$
  intervals in Fig.~\ref{fig-EWdist-L_UV}.  ($M_{\rm UV}$ is the same
  as the horizontal axis of Fig.~\ref{fig-Ando}.)  At the brightest
  $M_{\rm UV}$, LAEs are predominantly in the outflow phase, and their
  EWs are not larger than $\sim 100$~\AA, because their dust
  extinction is not large enough to EW enhancement by the geometrical
  effect.  On the other hand, the pre-outflow starbursts that are
  responsible for the large EW LAEs populate mainly the fainter
  $M_{\rm UV}$ range, because their UV luminosities are reduced by
  extinction.  A large amount of extinction is necessary to compensate
  the dust-independent factor $f_0$ and achieve large EW, and hence a
  trend of larger maximum EW values for smaller UV luminosities
  appears.  Therefore, the clumpy dust effect can explain not only the
  existence of large EW LAEs, but also the Ando effect quantitatively
  by the same physical process.

  It has been widely discussed that intrinsically larger galaxies have
  larger extinction on average.  For example, Adelberger \& Steidel et
  al. (2000) and Reddy et al. (2008) showed that mean extinction
  increases with the intrinsic (dust-corrected) UV luminosity for $z
  \sim 2-3$ LBGs, and Gawiser et al. (2006) indicated that
  IRAC/Spitzer non-detected (i.e., less massive) LAEs at $z = 3.1$
  have smaller dust amount than those detected by IRAC.  This trend
  may appear to be contradictory to our interpretation of the Ando
  effect, i.e., smaller extinction for UV brighter galaxies.  However,
  it should be noted that the Ando effect is in terms of the
  observable (i.e., dust extinction uncorrected) absolute magnitude of
  UV continuum, but not in terms of the intrinsic one.  There is a
  considerable scatter between the intrinsic and observable UV
  magnitudes because of the strong extinction in the rest-frame UV
  band.  As a result, we do not expect a clear trend between
  extinction and observed UV magnitude and hence it is not
  contradictory with the observations.  In fact, there are some
  observations that support our interpretation.  Reddy et al. (2008)
  and Buat et al. (2009) have reported that the maximum value of
  attenuation factor by dust increases toward fainter $M_\mathrm{UV}$
  (observable) while its mean value slightly decreases.  In our model,
  the large EW LAEs are produced by strong extinction with the clumpy
  ISM effect, resulting in larger maximum attenuation factor at
  fainter $M_\mathrm{UV}$.  However, the dominant LAEs in number
  density in this magnitude range are, in fact, quiescent galaxies
  with small EW value, as seen in Fig.~\ref{fig-EWdist-L_UV}, and
  hence the mean attenuation factor does not increase toward fainter
  $M_\mathrm{UV}$.

%%%%%%%%%%%%%%%%%%%%%%%%%%%%%%%%%%%%%%%%%%%%%%%%%%%%%%%%%%%%%%%%%%%%%  
 \section{DISCUSSION}
 \label{section:discussion}

  \subsection{LAEs at $z\gtrsim 6$ and IGM Transparency: 
  Implication for the Cosmic Reionization}
  \label{section:reionization}
  
  Although the $\lya$ LF predictions by the KTN07 model are in
  reasonable agreement with the observations at $z =3$--6, it becomes
  discrepant rather suddenly with the observed data at $z \gtrsim 6$.
  KTN07 argued that this discrepancy can be resolved if the IGM
  transmission for $\lya$ photons, $\tlya$ (the fraction of $\lya$
  photons that can be transmitted without absorption by neutral
  hydrogen in the IGM), becomes small at $z \gtrsim 6$.  A value of
  $\tlya \sim 0.5$--0.6 was inferred in order to make our prediction
  consistent with the observed LAE $\lya$ LFs at $z=6.56$ by Kashikawa
  et al. (2006) and at $z = 6.96$ by Iye et al. (2006) and Ota et al.
  (2008).  (Here, $\tlya$ is defined as a relative transmission
  compared with that at $z \lesssim 6$, and hence $\tlya = 1$ at $z
  \lesssim 6$.  See \S~\ref{section:model-LyA-escape}.)  If this
  interpretation is correct, it indicates a significantly higher
  neutral fraction in IGM at $z \gtrsim 6$ than lower redshifts, and
  we may be observing the end of the cosmic reionization, although
  quantitative conversion from $\tlya$ to IGM neutral fraction $x_{\rm
  HI} \equiv n_{\rm HI}/n_{\rm H}$ is rather uncertain (Santos 2004;
  Haiman \& Cen 2005; Dijkstra et al. 2007a, b).  Here, we examine
  whether the other statistical quantities of UV LFs and EW
  distributions are quantitatively consistent with the above
  interpretation.
  
\begin{figure}
 \epsscale{.6}
 \plotone{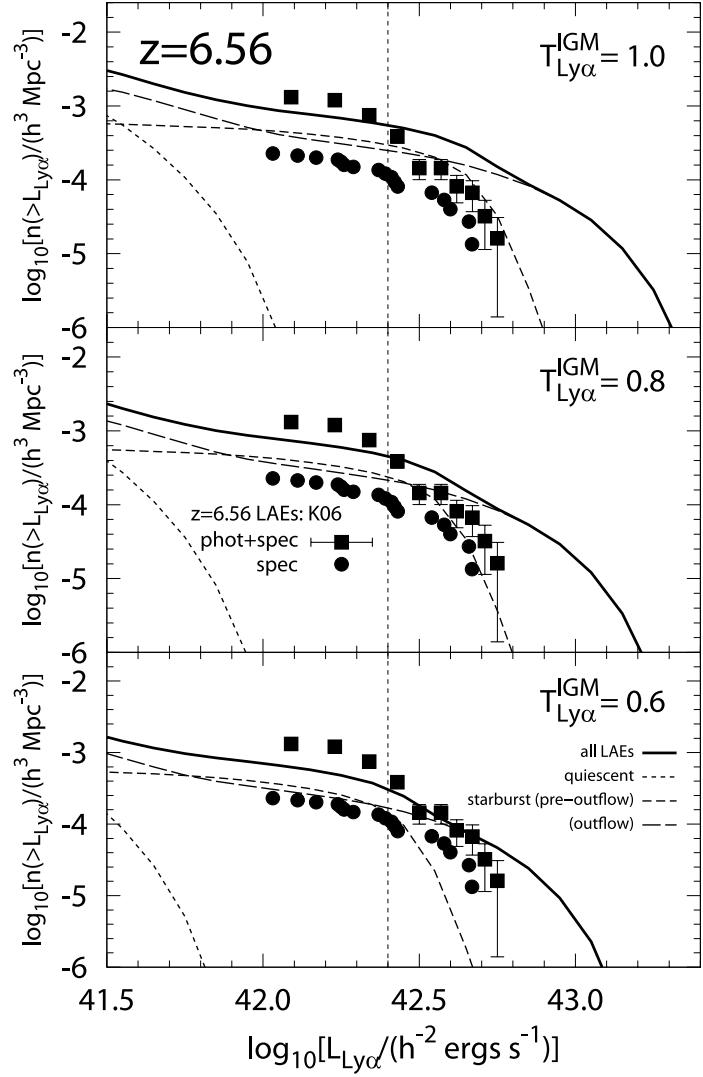}
 \caption{ 
   Comparisons between the cumulative $\lya$ LF data of Kashikawa et
   al. (2006) at $z=6.56$ and model predictions with three different
   values of the IGM transmission for $\lya$ photons, $\tlya =$1.0
   (top), 0.8 (middle), and 0.6 (bottom).  The thick solid curve is
   for all model LAEs, and the other curves are for model LAEs with
   different phases (the same line markings as those of the lower
   panels of Fig.~\ref{fig-LyALF-LAE}).  The filled squares are all
   LAE candidates in the K06 sample, while the filled circles are
   those with spectroscopic confirmation.  The vertical dashed line
   indicates the threshold $\lya$ luminosity of $10^{42.4}\ h^{-2}\
   \mathrm{ergs\ s^{-1}}$, which will be applied in the comparisons
   with UV LF, EW distribution, and $M_\mathrm{UV}-\ewlya$ plane (see
   Fig.~\ref{fig-z6p6}).
 }
 \label{fig-z6p6-LyaLF}
\end{figure}

\begin{figure}
 \epsscale{1}
 \plotone{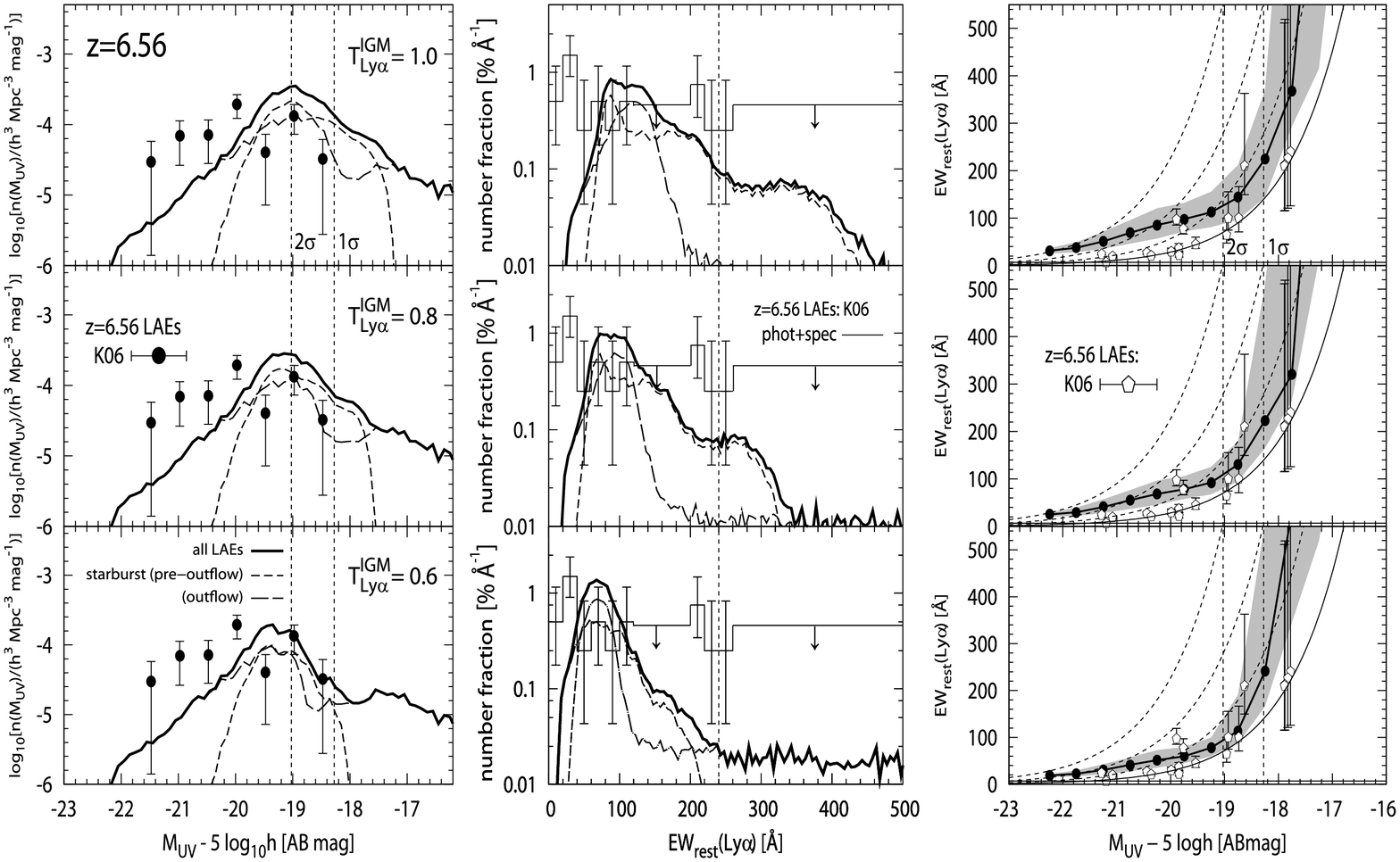}
 \caption{ 
   The same as Fig.~\ref{fig-z6p6-LyaLF}, but for comparisons with UV
   LF, EW distribution, and $M_\mathrm{UV}-\ewlya$ plane from the left
   to the right panels.  The line markings of the model curves in the
   left, middle, and right panels are the same as those of
   Figs.~\ref{fig-UVLF-LAE}, \ref{fig-EW}, and \ref{fig-Ando},
   respectively.  Here, only LAE candidates having $\lobs \ge
   10^{42.4}\ h^{-2}\ \mathrm{ergs\ s^{-1}}$ are used from the K06
   sample, and this condition is included in the model predictions as
   well.  The vertical dotted lines in the UV LF and
   $M_\mathrm{UV}-\ewlya$ plane indicate the 1 and 2~$\sigma$ limiting
   UV continuum magnitudes of Kashikawa et al. (2006).
 }
 \label{fig-z6p6}
\end{figure}

  Figure~\ref{fig-z6p6-LyaLF} shows the comparisons of $\lya$ LF
  between the model and observations at $z=6.56$, testing several
  different values of $\tlya$.  
  \placefigure{fig-z6p6} 
  This is identical to Fig.~4 in KTN07 but here for the updated
  outflow$+$dust (slab) model.  While $\tlya \sim 0.6$ is preferred in
  order to fit the bright end of the observational $\lya$ LF, model
  prediction is underestimated by a factor of $\sim 2$ at faint end
  compared with observation.  LAE candidates at such faint end have
  not yet been confirmed by spectroscopy, leaving room for possible
  contaminations (N. Kashikawa, private communication).  The
  uncertainty of the observational $\lya$ LF at faint end can be as
  large as a factor of $\sim 2$ because of the low detection
  completeness of $\sim 45$\% (Kashikawa et al. 2006).  Therefore, in
  order to avoid the uncertainty caused by contaminations and low
  completeness, we compare our model with the other statistical
  quantities (UV LF, EW distributions, and $M_\mathrm{UV}-\ewlya$
  plane) by using only photometric samples brighter than $\lobs \ge
  10^{42.4}\ h^{-2}\ \mathrm{ergs\ s^{-1}}$, at which the detection
  completeness is larger than $\sim 75$\%.

  The comparisons are presented in Figure~\ref{fig-z6p6}.  In our
  model prescription, the $\lya$ luminosities are simply reduced by
  the factor of $\tlya$, while $L_{\rm UV}$ is not changed.  As a
  consequence, the EW distribution is just shifted into the smaller EW
  direction with decreasing $\tlya$, and the positions of model
  galaxies in the $M_{\rm UV}$-EW plane simply move downward, while no
  change in UV LFs.  However, because of the threshold values of
  $\lth$ and $\ewth$, some model galaxies are excluded from the LAE
  selection as $\tlya$ decreased, resulting in a slight change in the
  faint end of UV LF.  This is simply because UV-faint galaxies are
  also $\lya$-faint on average, and hence they have a higher chance to
  be excluded from the sample by the threshold $\lya$ luminosity.
  
  The EW distribution at $\ewlya \lesssim 120$~{\AA} shows a better
  agreement with the model when $\tlya \sim 0.6$ than the case of
  $\tlya = 1$.  On the other hand, the number fraction of LAEs with
  $\ewlya \sim 200-260$~{\AA} seems to favor $\tlya = 1$.  However, it
  should be noted that the EW measurement is quite uncertain for large
  EW LAEs, because they have faint UV luminosity.  This can be clearly
  seen in the $M_\mathrm{UV}-$EW planes (the right panels of
  Fig.~\ref{fig-z6p6}), where we show the vertical dotted lines
  indicating $1\sigma$ and $2\sigma$ lines of the signal-to-noise of
  UV luminosity measurement.  All LAEs with $\ewlya > 200$~{\AA} in
  the K06 sample are less than $2\sigma$ about UV luminosity, and the
  errors of $\ewlya$ are quite large, as shown in this plot.
  Therefore, we consider that the EW distribution at $\ewlya >
  200$~{\AA} cannot reliably be compared with the model.  When we
  concentrate on the reliable regions of $\ewlya < 200$~{\AA} in the
  EW distribution and $M_{\rm UV} - 5 \log h < -19.03$ (2$\sigma$
  line) in $M_\mathrm{UV}-$EW plane, we find that the model with
  $\tlya \sim 0.6$ gives better fits than that with $\tlya = 1$.
  Therefore we conclude that the indication of $\tlya \sim 0.6$
  obtained by KTN07 is quantitatively strengthened by these
  comparisons.  It should also be noted that the comparison in UV LF
  also favors $\tlya \sim 0.6$, by the threshold effect about $\lobs$.

  \subsection{Redshift Evolution of LAE UV LF and Selection Effects} 
  \label{subsec:chUV}
  
\begin{figure}
 \epsscale{1}
 \plotone{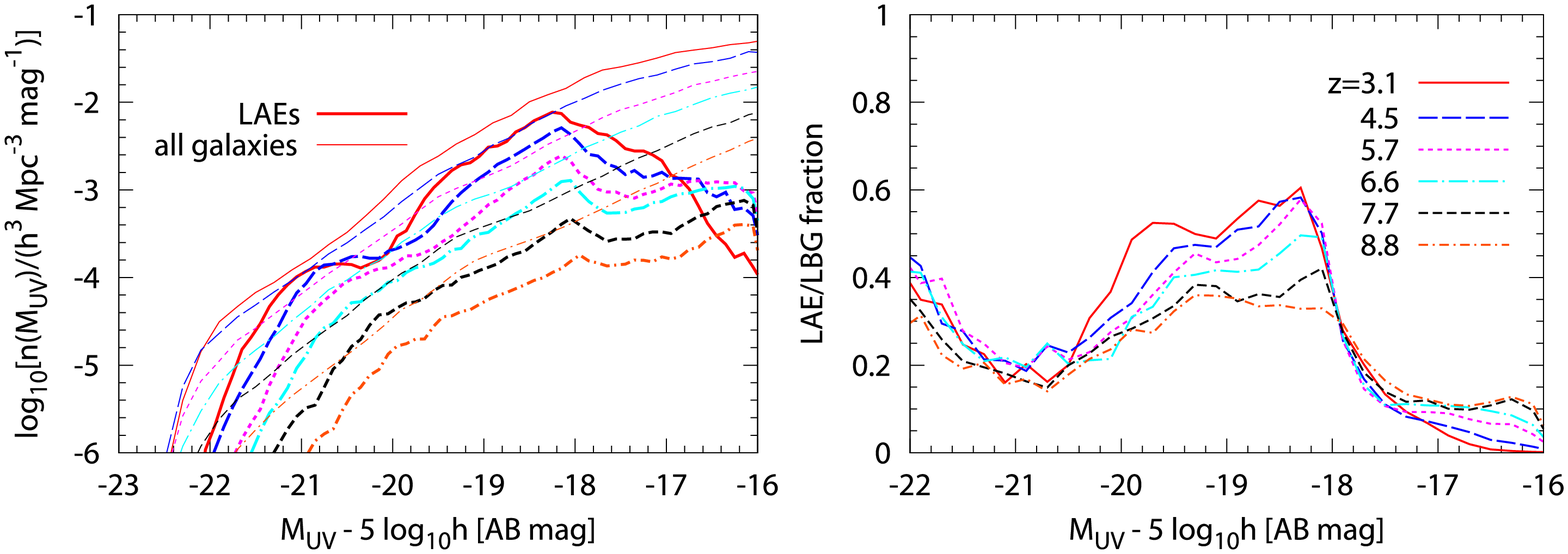}
 \caption{ 
   Redshift evolution of LAE UV LF (\textit{left}) and LAE fraction in
   LBGs as a function of UV luminosities (\textit{right}) predicted by
   our model from $z=3.1$ to $z=8.8$ with $\tlya =1$.  The dashed
   curves in the left panel show the UV LFs of all galaxies (or
   observationally, LBGs).  The galaxies with $\lobs \ge \lth
   =10^{41.5}~\unitlum$ and $\ewlya \ge \ewth =20$~{\AA} are selected
   as LAEs regardless of redshift.
 }
 \label{fig-UVLFev}
\end{figure}

  Recently, Ouchi et al. (2008) reported an ``anti-hierarchical''
  evolution of LAE UV LFs; they found that the LAE UV LFs obtained by
  the Subaru/XMM-Newton Deep Survey show increase of luminosity and/or
  number densities from $z=3.1$ to 5.7.  Moreover, several
  observational results indicate that the number fraction of LAEs in
  LBGs increases with redshift (e.g., Shimasaku et al. 2006; Ouchi et
  al. 2008; Shioya et al. 2009). This trend has been theoretically
  quantified by Samui et al. (2009) as the LAE fraction of $\sim 0.1$
  and $\sim 1$ at $z=3-4$ and $z>5$, respectively, to fit their
  analytic model to the observational data of LAEs.  These trends are
  apparently in contradiction with our model prediction given in
  Fig.~\ref{fig-UVLFev}, where the UV LF of LAEs continuously decrease
  to higher redshifts beyond $z=3$, and the LAE fraction in LBGs is
  approximately constant against redshifts.  However, in order to
  discuss such redshift evolutions, it is important to carefully
  consider the dependencies of LAE LFs (both in terms of $\lya$ and UV
  continuum) and LAE fraction in LBGs on the threshold values of LAE
  selection: $\lth$ and $\ewth$.  In Fig.~\ref{fig-UVLFev}, we have
  assumed constant values for $\lth$ and $\ewth$ regardless of
  redshift.  (Note that the LAE fraction in LBGs is not a free
  parameter in our model, but it is derived from more physical
  modelings about $\fesclya$ taking also into account the adopted
  values of $\lth$ and $\ewth$ in an observed data set.)  Here we
  examine this issue and whether our model is consistent with the
  observations.

\begin{figure}
 \epsscale{1}
 \plotone{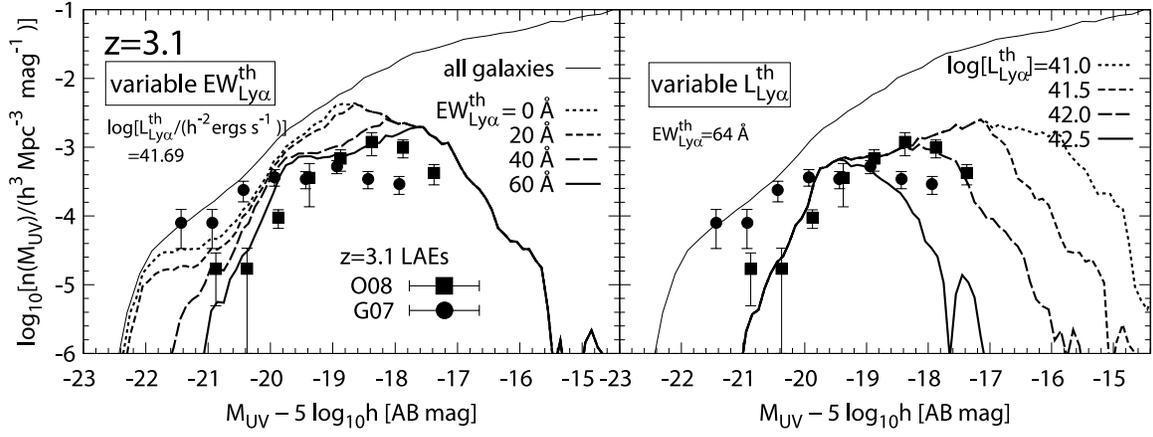}
 \caption{ 
   UV LFs of LAEs at $z=3.1$ predicted by our model with different
   values of threshold $\ewth$ (\textit{left}) and $\lth$
   (\textit{right}) in LAE selection.  The thin solid curve represents
   the UV LF of all galaxies including non-LAEs.  The observed data of
   UV LFs of LAEs at $z=3.1$ obtained by Gronwall et al. (2007, G07)
   and Ouchi et al. (2008, O08) are also shown.  Their LAE selection
   thresholds are $[\lth /(h^{-2}\ {\rm ergs\ s^{-1}}), \
   \ewth / \mathrm{\mathring{A}}] = (10^{41.79},\ 20)$ and
   $(10^{41.69},\ 64)$ for G07 and O08, respectively.
 }
 \label{fig-th}
\end{figure}

  The dependencies of LAE UV LF on the threshold values of $\ewth$ and
  $\lth$ predicted by our model at $z=3.1$ are shown in
  Figure~\ref{fig-th}.  It is found that difference of $\ewth$ results
  mainly in changes of the bright end of UV LFs, because the UV
  brightest LAEs have lower EWs on average by the Ando effect and
  hence they are more easily affected by $\ewth$.  On the other hand,
  $\lth$ affects mainly the faint end of UV LFs, simply because
  UV-faint LAEs have small $\lya$ luminosities on average although
  there is a considerable scatter between $\lya$ and UV luminosities.
  In Figure~\ref{fig-th}, the observed UV LFs of the LAEs obtained by
  Gronwall et al. (2007) and Ouchi et al. (2008) at the same redshift
  but with different LAE selection thresholds are also shown.  The
  behavior of the bright end of the observed UV LFs against $\ewth$ is
  quantitatively well reproduced by our model.  The faint-end behavior
  against $\lth$ is also consistent with the model prediction, at
  least qualitatively.

  Our model indicates that the observed anti-hierarchical evolution of
  LAE UV LF reported by Ouchi et al. (2008) is merely a consequence of
  a selection effect: decreasing $\ewth$ with increasing redshift.
  These results demonstrate the importance of the selection conditions
  of LAEs when the redshift evolutions of their statistical quantities
  are discussed.

%%%%%%%%%%%%%%%%%%%%%%%%%%%%%%%%%%%%%%%%%%%%%%%%%%%%%%%%%%%%%%%%%%%%%
 \section{Conclusions}
 \label{section:conclusions}

 We have performed a comprehensive comparison between a theoretical
 model of high-$z$ LAEs in the framework of hierarchical structure
 formation and a variety of observations about statistical quantities
 such as UV continuum luminosity function (UV LF), equivalent width
 distributions, and correlation between UV luminosity and EW at
 redshifts $z \sim 3$--6.  The model used in this work was constructed
 by our previous study (KTN07), showing a good agreement with the
 observed $\lya$ LF of LAEs.  All the model parameters about LAEs have
 already been determined by fitting to the observed $\lya$ LF at
 $z=5.7$ in KTN07, and there is no more free parameters in the new
 comparison made in this work, giving an objective test for the
 validity of our LAE model.  In the comparisons between the model and
 observations, we paid a particular attention about the selection
 conditions of LAEs; the theoretical model predictions are compared
 separately with different observations even at the same redshift, by
 adopting the same conditions as those in each observation for the
 model predictions.

 In our model, extinctions by dust for $\lya$ photons and continuum
 photons are treated separately, and the best-fit parameters to $\lya$
 LF obtained by KTN07 indicate that the extinction of $\lya$ photons
 is significantly less effective than that for continuum photons
 around the $\lya$ wavelength.  This suggests that the dust geometry
 in ISM of most LAEs is clumpy, and this is consistent
 with recent independent observational studies by Finkelstein et al.
 (2008, 2009a, 2009b).  This clumpy dust distribution plays an
 important role to reproduce the observations of the statistical
 quantities of LAEs considered in this paper.

 We found that the predicted statistical quantities are in nice
 agreement with various observational data in the redshift range of
 $z\sim 3-6$.  In particular, our model naturally reproduces (1) the
 existence of large EW LAEs ($\gtrsim 240$~\AA) without invoking Pop
 III stars or top-heavy IMF, and (2) the Ando effect, which is the
 observed trend of smaller mean $\lya$ EW for more UV-luminous LAEs.
 We carefully examined the physical origin of these results, and found
 that both the stellar population and clumpy dust are important to
 reproduce large EW LAEs.  EW becomes large when the stellar
 population is young and/or metal-poor, but these effects are not
 sufficient to produce LAEs with EW $\gtrsim 240$~{\AA} only by these.
 EW is further enhanced by clumpy dust in LAEs with large extinction,
 resulting in large EW LAEs.  Our model then predicts that there is a
 correlation or trend that LAEs with larger EWs have larger reddening,
 and we found that this prediction is quantitatively consistent with
 the recent observational results.  The Ando effect is again explained
 by the clumpy dust effect, because large EW LAEs need a significant
 amount of extinction, and such galaxies have fainter UV luminosities
 due to the extinction.  Therefore, we could explain all the
 statistical quantities of LAEs under the standard scenario of
 hierarchical galaxy formation, by normal stellar populations without
 Pop III stars or top heavy IMF, but requiring clumpy dust
 distribution in ISM of LAEs.
 
 For LAEs at $z\gtrsim 6$, the observational data prefer the model
 predictions with a smaller $\tlya$ than that at $z \lesssim 6$, where
 $\tlya$ is the IGM transmission for $\lya$ photons.  The result is
 consistent with that of our previous study (KTN07) in terms of $\lya$
 LFs, but here we confirmed this result by more statistical quantities
 (UV LF, EW distributions, and $M_{\rm UV}$-$\ewlya$ plane).
 Therefore, this result provides a further evidence that $\tlya$ is
 rapidly decreasing with redshift beyond $z \sim 6$, giving an
 important implication for the cosmic reionization.

 The dependence of LAE UV LF and LAE fraction in LBG population on the
 selection criteria of LAEs about $\lya$ luminosity and EW was also
 discussed.  We have shown that an apparent evolutionary effect can be
 observed if one uses different selection criteria for observations at
 different redshifts.  It is quite important that the same selection
 criteria are applied to the observational data in order to discuss
 the redshift evolution of LAEs without observational bias.

 There are a lot of theoretical uncertainties in the physics of LAEs,
 and our model may not be unique to explain the observed data.
 However, we emphasize that our work is the first to compare a
 theoretical model of LAEs to almost all of available observational
 quantities (LAE LFs in terms of $\lya$ and UV continuum, EW
 distribution, EW-$M_{\rm UV}$ relation, etc.) at various redshifts
 with a consistent set of model parameters and with the LAE selection
 criteria in each observation appropriately taken into account.  It
 seems that extinction by the clumpy ISM is a unique and necessary
 ingredient to reproduce all of the observed data within the framework
 our model, if we do not consider somewhat exotic assumptions such as
 the contribution from Pop III stars or top-heavy IMF.  Of course, our
 model does not exclude the possibilities of Pop III or top-heavy IMF,
 and we need more data and theoretical studies to discriminate these
 possibilities.  However, we have shown that the standard picture of
 galaxy formation within the framework of hierarchical structure
 formation can be wholly consistent with the available LAE data,
 simply by introducing the clumpy ISM effect for extinction of $\lya$
 photons by dust.
 
 Our theoretical model for the various observational quantities of
 LAEs at various redshifts would be helpful in planning an LAE survey
 at even higher redshifts and interpreting such data sets in future
 studies.  The numerical data on these quantities of LAEs are
 available upon request to the authors.

%%%%%%%%%%%%%%%%%%%%%%%%%%%%%%%%%%%%%%%%%%%%%%%%%%%%%%%%%%%%%%%%%%%%%
 
\acknowledgments

We would like to thank the referee for useful comments.  We would also
like to thank Steven Finkelstein, Caryl Gronwall, Nobunari Kashikawa,
Masami Ouchi, and Kazuhiro Shimasaku for providing their observational
data and for useful discussions.  The numerical calculations were in
part carried out on SGI Altix3700 BX2 at Yukawa Institute for
Theoretical Physics of Kyoto University.  This research was supported
by the Japan Society for the Promotion of Science (JSPS) through the
Grant-in-Aid for Scientific Research (19740099 and 19035005 for TT,
18749007 for MN).  MARK was supported by the Research Fellowship for
Young Scientists from the JSPS.  TT appreciates the support by the
Global COE Program ``The Next Generation of Physics, Spun from
Universality and Emergence'' from the Ministry of Education, Culture,
Sports, Science and Technology (MEXT) of Japan.

%%%%%%%%%%%%%%%%%%%%%%%%%%%%%%%%%%%%%%%%%%%%%%%%%%%%%%%%%%%%%%%%%%%%%
 
% For papers with more than 8 authors, the last name and initials of the
% first author only should be listed, followed by a comma and et al. 
% The format in [] should be [Author(Year)]. If ``(Year)'' is forgotten
%  to write, the commands of \citet and \citep do not work
%  appropriately; in the wrong cases, only the number is displayed
%  and the author names and published year do not shown.

%\clearpage
 
%%%%%%%%%%%%  Tables  %%%%%%%%%%%%%%%%%%%%%%%%%%%%%%%%%%%%%%%
 \begin{deluxetable}{cccccc}
  \tabletypesize{\scriptsize}
  \tablewidth{0pt}
  \tablecolumns{6}
  \tablecaption{Compilation of Observational Criteria for $\lya$ EW
  and Line Luminosity}

  \tablehead{
  \colhead{$z$} &
  \colhead{Ref.} &
  \colhead{$\ewth$ [\AA]} &
  \colhead{$\log_{10}{[\lth/(\unitlum)]}$} &
  \colhead{$N_\mathrm{phot}$} &
  \colhead{$V_\mathrm{eff}~[10^4~ h^{-3}~ \mathrm{Mpc^3}]$}
  }

  \startdata
  3.1............  & G07      & 20 & 41.79 & 160 & 3.8\\
                   & O08      & 64 & 41.69 & 356 & 24.0\\
  3.7............  & O08      & 44 & 42.29 & 101 & 23.0\\
  4.5............  & D07      & 14 & 42.29 & $\sim 350$ & 50.7\\
  5.7............  & S06      & 20 & 42.04 & 89  & 6.2\\
                   & O08      & 27 & 42.17 & 401 & 31.6\\
  6.56............ & T05, K06 &  7 & 41.99 & 58  & 7.4
  \enddata

  \tablecomments{
  Col. (1): Redshift. 
  Col. (2): References. T05: Taniguchi et al. (2005), K06: Kashikawa
  et al. (2006), S06: Shimasaku et al. (2006), D07: Dawson et
  al. (2007), G07: Gronwall et al. (2007), O08: Ouchi et al. (2008). 
  Cols. (3) and (4): Threshold values of rest-frame $\ewlya$ and
  $\lya$ line luminosity, respectively, to select LAEs. 
  Col. (5): Number of photometrically selected LAE candidates. 
  Col. (6): Effective survey volume.
  }

  \label{tab-comp}
 \end{deluxetable}

%%%%%%%%%%%%%% Figures %%%%%%%%%%%%%%%

\end{document}